\documentclass[twocolumn]{aastex61}
\usepackage{color}

\usepackage{natbib}
\begin{document}

\title{44 New \& Known M dwarf Multiples in the SDSS-III/APOGEE M Dwarf Ancillary Science Sample}








\author{Jacob Skinner}
\affil{Department of Physics \& Astronomy, Western Washington University, Bellingham, WA 98225, USA}
\email{skinnej3@wwu.edu}

\author{Kevin R. Covey}
\affil{Dept. of Physics \& Astronomy, Western Washington University, Bellingham, WA 98225, USA}
\email{kevin.covey@wwu.edu}

\author{Chad F. Bender}
\affil{Department of Astronomy and Steward Observatory, University of Arizona, Tucson, AZ 85721, USA}
\email{cbender@email.arizona.edu}

\author{Noah Rivera}
\affil{Department of Astronomy and Steward Observatory, University of Arizona, Tucson, AZ 85721, USA}

\author{Nathan De Lee}
\affil{Department of Physics, Geology, and Engineering Technology, Northern Kentucky University, Highland Heights, KY 41099, USA}
\affil{Department of Physics \& Astronomy, Vanderbilt University, Nashville, TN 37235, USA}

\author{Diogo Souto}
\affil{Observat\'orio Nacional, Rua General Jos\'e Cristino, 77, 20921-400 S\~ao Crist\'ov\~ao, Rio de Janeiro, RJ, Brazil}

\author{Drew Chojnowski}
\affil{Apache Point Observatory and New Mexico State University, P.O. Box 59, Sunspot, NM 88349-0059, USA}

\author{Nicholas Troup}
\affil{Department of Physics, Salisbury University, 1101 Camden Ave, Salisbury, MD 21801, USA}


\author{Carles Badenes}
\affil{Department of Physics and Astronomy, University of Pittsburgh, Allen Hall, 3941 O'Hara St, Pittsburgh PA 15260, USA}

\author{Dmitry Bizyaev}
\affil{Apache Point Observatory and New Mexico State University, P.O. Box 59, Sunspot, NM 88349-0059, USA}

\author{Cullen H. Blake}
\affil{Department of Physics and Astronomy, University of Pennsylvania, 209 South 33rd Street, Philadelphia, PA 19104, USA}

\author{Adam Burgasser} 
\affil{Center for Astrophysics and Space Science, University of California San Diego, La Jolla, CA 92093, USA}

\author{Caleb Ca\~nas}
\affil{Department of Astronomy \& Astrophysics, Pennsylvania State, 525 Davey Lab, University Park, PA 16802, USA}

\author{Joleen Carlberg}
\affil{Space Telescope Science Institute, Baltimore, MD, 21218, USA}

\author{Yilen G\'omez Maqueo Chew}
\affil{Instituto de Astronom\'ia, Universidad Nacional Aut\'onoma de M\'exico,  Ciudad Universitaria, Ciudad de M\'exico, 04510, M\'exico} 

\author{Rohit Deshpande}
\affil{Department of Astronomy \& Astrophysics, Pennsylvania State, 525 Davey Lab, University Park, PA 16802, USA}

\author{Scott W. Fleming}
\affil{Space Telescope Science Institute, Baltimore, MD, 21218, USA}

\author{J. G. Fern\'andez-Trincado}
\affil{Departamento de Astronom\'ia, Casilla 160-C, Universidad de Concepci\'on, Concepci\'on, Chile}
\affil{Institut Utinam, CNRS UMR6213, Univ. Bourgogne Franche-Comt\'e, OSU THETA, Observatoire de Besan\c{c}on, BP 1615, 25010 Besan\c{c}on Cedex, France}

\author{D. A. Garc\'ia-Hern\'andez} 
\affil{Instituto de Astrof\'isica de Canarias (IAC), V\'ia Lactea s/n, E-38205 La Laguna, Tenerife, Spain}
\affil{Departamento de Astrof\'isica, Universidad de La Laguna (ULL), E-38206 La Laguna, Tenerife, Spain}

\author{Fred Hearty}
\affil{Department of Astronomy \& Astrophysics, Pennsylvania State, 525 Davey Lab, University Park, PA 16802, USA}

\author{Marina Kounkel}
\affil{Dept. of Physics \& Astronomy, Western Washington University, Bellingham, WA 98225, USA}

\author{Pen\'elope Longa-Pe\~ne}
\affil{Unidad de Astronom\'ia, Facultad de Ciencias B\'asicas, Avenida Angamos 601, Antofagasta 1270300, Chile}

\author{Suvrath Mahadevan}
\affil{Department of Astronomy \& Astrophysics, Pennsylvania State, 525 Davey Lab, University Park, PA 16802, USA}

\author{Steven R. Majewski}
\affil{Department of Astronomy, University of Virginia, Charlottesville, VA 22904-4325, USA}

\author{Dante Minniti}
\affil{Pontificia Universidad Cat\'olica de Chile, Instituto de Astrofısica, Av. Vicuna Mackenna 4860, 782-0436 Macul, Santiago, Chile}

\author{David Nidever}
\affil{Department of Physics, Montana State University, Bozeman, MT 59717, USA}

\author{Audrey Oravetz}
\affil{Apache Point Observatory and New Mexico State University, P.O. Box 59, Sunspot, NM 88349-0059, USA}

\author{Kaike Pan}
\affil{Apache Point Observatory and New Mexico State University, P.O. Box 59, Sunspot, NM 88349-0059, USA}

\author{Keivan Stassun}
\affil{Department of Physics and Astronomy, Vanderbilt University, VU Station 1807, Nashville, TN 37235, USA}

\author{Ryan Terrien}
\affil{Department of Physics \& Astronomy, Carleton College, Northfield MN, 55057, USA}

\author{Olga Zamora}
\affil{Instituto de Astrof\'isica de Canarias (IAC), V\'ia Lactea s/n, E-38205 La Laguna, Tenerife, Spain}
\affil{Departamento de Astrof\'isica, Universidad de La Laguna (ULL), E-38206 La Laguna, Tenerife, Spain}


\begin{abstract}
	Binary stars make up a significant portion of all stellar systems. Consequently, an understanding of the bulk properties of binary stars is necessary for a full picture of star formation. Binary surveys indicate that both multiplicity fraction and typical orbital separation increase as functions of primary mass. Correlations with higher order architectural parameters such as mass ratio are less well constrained.
	We seek to identify and characterize double-lined spectroscopic binaries (SB2s) among the 1350 M dwarf ancillary science targets with APOGEE spectra in the SDSS-III Data Release 13.
	We measure the degree of asymmetry in the APOGEE pipeline cross-correlation functions (CCFs), and use those metrics to identify a sample of 44 high-likelihood candidate SB2s. At least 11 of these SB2s are known, having been previously identified by Deshapnde et al, and/or El Badry et al. We are able to extract radial velocities (RVs) for the components of 36 of these systems from their CCFs. With these RVs, we measure mass ratios for 29 SB2s and 5 SB3s. We use Bayesian techniques to fit maximum likelihood (but still preliminary) orbits for 4 SB2s with 8 or more distinct APOGEE observations.
	The observed (but incomplete) mass ratio distribution of this sample rises quickly towards unity. Two-sided Kolmogorov-Smirnov tests find probabilities of 18.3\% and 18.7\%, demonstrating that the mass ratio distribution of our sample is consistent with those measured by Pourbaix et al. and Fernandez et al., respectively.
\end{abstract}


\section{INTRODUCTION}
Models of stellar formation and evolution make predictions about the distribution and frequency of stellar binaries. Fragmentation of a protostellar core or circumstellar disk can produce the requisite pair of pre-main sequence stars \citep[e.g., ][]{Offner2010}, but only at much larger separations \citep[$\sim$100-1000$+$ AU;][]{Tohline2002, Kratter2011} than those that characterize close\footnote{with separations on the order of $\bm{\sim}$1AU. i.e. Non-interacting and spectroscopic.} binaries. Dynamical processes presumably drive some of these wider binaries into a close configuration, but the nature and timescale of this evolution remains unclear: mechanisms that may play a role include dynamical friction from gas in the surrounding disk or core \citep[e.g., ][]{Gorti1996}, gravitational interactions in/dynamical decay of few-body systems \citep[e.g., ][]{Reipurth2001, Bate2002}, Kozai oscillations \citep{Fabrycky2007} and/or tidal friction \citep[e.g., ][]{Kiseleva1998}.

Empirical study has provided some data with which to test these models. 
The multiplicity fraction (MF, $\frac{\# multiples}{all\,stars}$) is known to be an increasing function of primary mass: the lowest multiplicity rates are observed for substellar systems (MF $11^{+7}_{-2}\%$ implying a companion fraction (CF) $\frac{\# stars\,with\,companions}{all\,stars}$of $\approx 20\%$; \citet{Burgasser2006}), and rise into the M dwarf regime, where the seminal measurement of the companion fraction over all separations remains that of \citet{FischerMarcy1992}: $42\% \pm 9\%$ for separations of 0.04-10$^4$ AU.  Yet larger multiplicity rates are found for stars of G-type ($46\% \pm 2\%$; \citet{Raghavan2010}) and F-type and earlier ($100\% \pm 20\%$; \citet{Duchene2013}). There is also mounting evidence of a trend of binary separation increasing with primary mass \citep{Ward-Duong2015}. When corrected for incompleteness, the mass ratio distribution of close binaries is mostly flat \citep{Moe&DiStefano2017}.
\par
M dwarfs are a particularly common result of the star formation process, and by virtue of their low masses, provide leverage for probing the link between primary mass, companion fraction, and orbital separation. Since the survey of \citet{FischerMarcy1992}, additional M dwarf multiplicity surveys have been conducted by \citet{Clark2012}, \citet{Shan2015} and \citet{Ward-Duong2015}, who used various observational techniques to identify 22, 12 and 65 multiple systems within samples of 1452, 150, and 245 M dwarfs, respectively.  These measurements are consistent with the \citet{FischerMarcy1992} result, suggesting a CF of 26-35\% for separations outside a key gap in coverage from 0.4-3 AU.
The near-infrared spectra of the APOGEE survey are well suited to detect the faint, cool companions of M dwarfs. This gives us a window into the dynamic evolution of early systems, as well as developed systems in the low period regime. A survey of M dwarf double-lined spectroscopic binaries (SB2s) in clusters and the field could detect changes in the close binary fraction with age, providing a valuable clue as to whether low period binaries most often mutually form up close, or evolve through 3 body dynamics with a 3rd, distant companion. 
\par
In this paper, we search the APOGEE spectroscopic database for close, double-lined spectroscopic binaries with low-mass, M dwarf primaries. We utilize a classic approach, searching for sources whose spectra include two or more sets of photospheric absorption lines, with a clear radial velocity offset in at least one APOGEE observation.  This approach compliments the recent search conducted by \citet{ElBadry2017}, using the direct spectral modeling approach validated by \citet{ElBadry2018}. The search completed by \citet{ElBadry2017} is sensitive to multiple systems over a much larger range of orbital separations, as their method can detect spectral superpositions even with no radial velocity offset. While their search is sensitive to a much broader range of parameter space in the dimension of orbital separation, their spectral modeling approach is limited to stars with  T$_{eff} >$ 4000 K, providing motivation for a directed search for close, low-mass spectroscopic binaries. 

We begin by introducing the observational data and describe our sample selection in \S 2. We describe our data analysis procedure, mass ratio measurements, and mass ratio distribution in \S 3. Section \S 4 contains the description of our orbit fitting procedure and results for 4 targets. Finally, we present our results in \S 5 and summarize our conclusions in \S 6. Appendix A contains notes on a mass estimation calculation mentioned in \S 4.2.

\section{OBSERVATIONS AND SAMPLE SELECTION}

\subsection{SDSS-III APOGEE M dwarf Ancillary Targets}
    The SDSS-III \citep{Eisenstein2011} APOGEE M dwarf Ancillary Program \citep{Deshpande2013, Holtzman2015} was designed to produce a large, homogeneous spectral library and kinematic catalog of nearby low-mass stars; these data products are useful for investigations of stellar astrophysics \cite[e.g.][]{Souto2017, Gilhool2018}, and for refining targeting procedures for current and future exoplanet search programs. These science goals are uniquely enabled by the APOGEE spectrograph \citep{Wilson2010, Wilson2012}, which acquires high resolution (R$\sim$22,000) near-infrared spectra from each of 300 optical fibers.  As deployed at the 2.5 meter SDSS telescope \citep{Gunn2006}, the APOGEE spectrograph achieves a field-of-view with a diameter of 3 degrees, making it a highly efficient instrument for surveying the stellar parameters of the constituents of Galactic stellar populations \citep{Majewski2017}. The SDSS DR13 data release \citep{Albareti2017} includes 7152 APOGEE spectra of 1350 stars targeted by this ancillary program. Methods used to select targets for the SDSS ancillary program are described in full by \citet{Deshpande2013} and \citet{Zasowski2013}; briefly, the targets were selected with one of the following methods: 
   
\begin{itemize}
\item{ stars of spectral type M4 or later, typically toward the fainter end of APOGEE's sensitivity range (H $\gtrsim$ 10), were targeted by applying a set of magnitude (7 $< H <$12) and color ($V - K_s >$ 5.0; 0.4 $< J - H < $0.65; 0.1 $< H - K_s <$ 0.42) cuts to the catalog of northern high proper motion ($\mu >$ 150 mas yr$^{-1}$) stars assembled by \citet{Lepine2005}.}
\item{M dwarfs of all spectral sub-classes, typically toward the brighter end of APOGEE's sensitivity range (H $\lesssim$ 10), were identified by applying simple spatial (DEC $>$ 0) and magnitude (H $>$ 7) cuts to the all-sky catalog of bright M dwarfs assembled by \citet{Lepine2011}.}
\item{calibrators with precise, stable radial velocities (as measured by the California Planet Search team), reliable measurements of rotation velocity \citep[$v$ sin $i$; ]{Jenkins2009}, active M dwarfs in the Kepler field \citep{Ciardi2011,Walkowicz2011}, or targets in the input catalog of the MEarth Project \citep{Nutzman2008} were individually added to the sample.}
\end{itemize}
    
\begin{figure}[t!]
\centerline{\includegraphics[scale=.45]{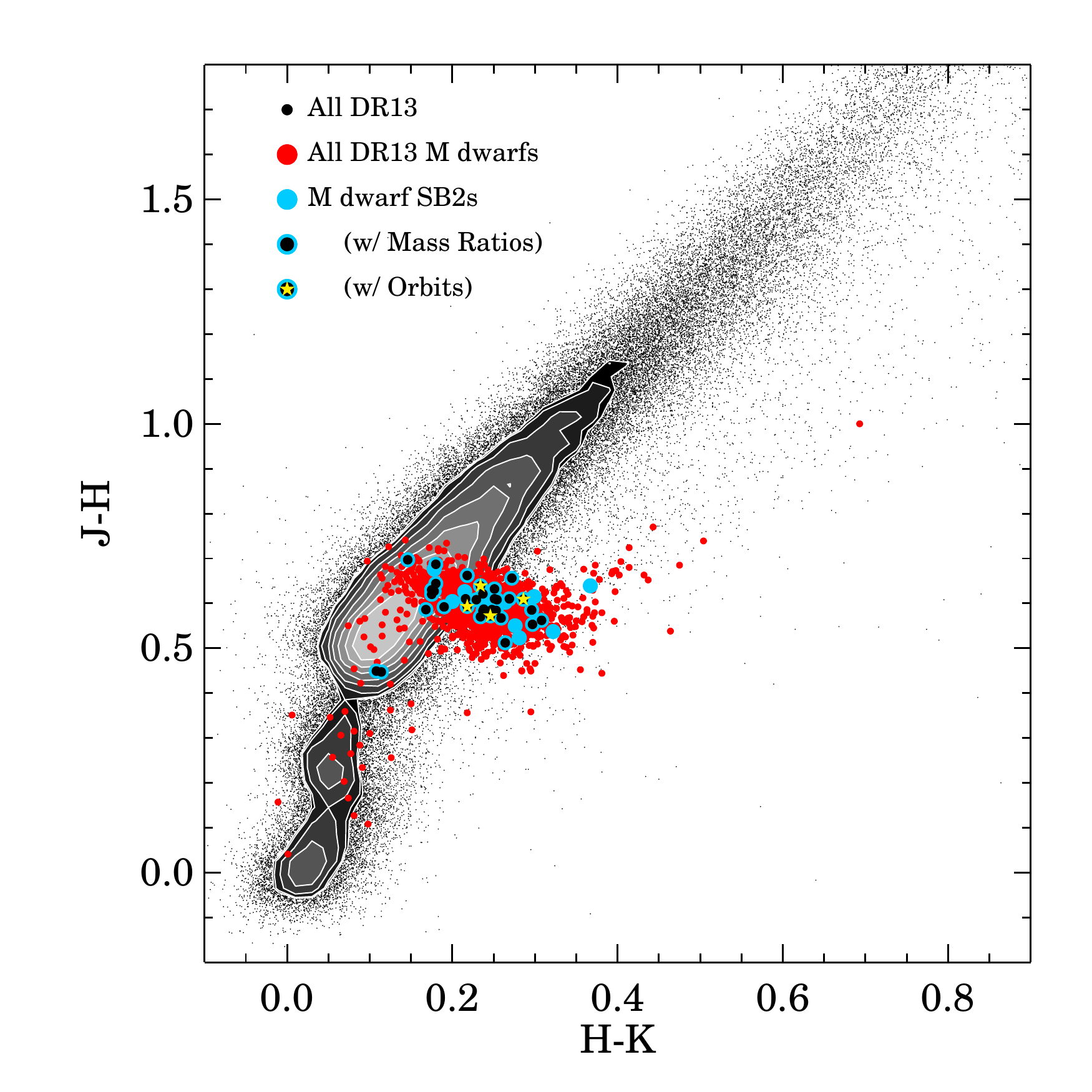}}
\caption{$J-H$ vs. $H-K_s$ color-color diagram of DR13 APOGEE targets.  The full DR13 sample is shown as small points, and grayscale contours in areas of color-space where individual points can no longer be distinguished. M dwarf ancillary targets are shown as solid red dots, demonstrating the clear divergence from the reddened giant branch which makes up the bulk of the APOGEE dataset. Candidate SB2s are indicated with cyan dots; sources for which we infer mass ratios and full orbital fits are highlighted with a black central dot and surrounding ring, respectively.}
\label{fig:JHK_colorcolor}
\end{figure}

\begin{figure}[t!]
\centerline{\includegraphics[scale=.50]{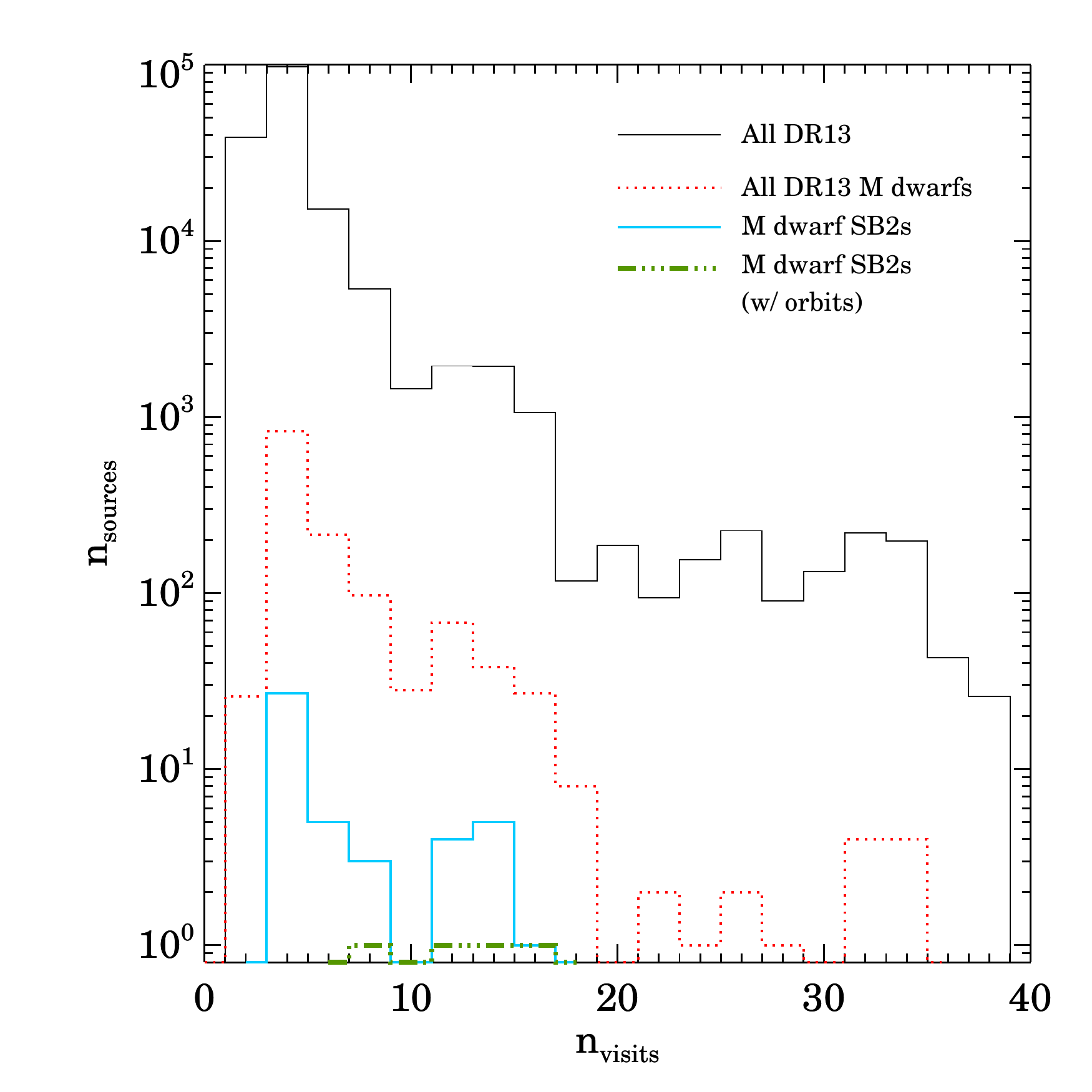}}
\caption{Histograms of the number of visits observed for different classes of APOGEE DR13 targets. The APOGEE M dwarf sample exhibits the same overall distribution of visits as the rest of the survey targets; the M dwarf SB2 candidates are modestly biased towards a larger number of visits, with $\sim$1 more visit per system in both the median and the mean than the broader DR13 sample.
}
\label{fig:Mdwarf_visits}
\end{figure}

Figure \ref{fig:JHK_colorcolor} shows the location of these 1350 ancillary targets in $J-H$ vs. $H-K_s$ color-color space, along with the full DR13 sample shown for context. Figure \ref{fig:Mdwarf_visits} compares the number of APOGEE observations obtained for objects identified here as binaries, relative to the number of observations obtained for the full DR13 sample and the subset of M dwarf ancillary science targets. On average, sources identified as SB2s have one more APOGEE observation than the median for the M dwarf ancillary science sample, reflecting the advantage that multi-epoch observations provide for identifying RV variable sources.
    
    \subsection{\textit{Identification as SB2s} } 
     
     Candidate SB2s were identified with an approach similar to that of \citet{Fernandez2017} (F17), who flagged sources with significant asymmetries in the cross-correlation functions (CCFs) calculated by the APOGEE pipeline \citep{Nidever2015, GarciaPerez2016, Grieves2017}. Following \citet{Fernandez2017}, we characterized the asymmetry in each CCF using the R parameter originally developed by \citet{Tonry1979}:
     $$R=\frac{H}{\sqrt{2}\sigma_a}$$
where H is the maximum of the CCF, and $\sigma_a$ is the RMS of the anti-symmetric portion of the CCF. In this formalism, lower R values indicate sources with larger asymmetries in their CCF functions. To better identify sources with CCF asymmetries at physically meaningful velocity separations, we computed distinct R values for windows of differing widths around each CCF's central peak. Specifically, we computed R values for the central 51, 101, and 151 lags in each CCF, which we denote as R$_{51}$, R$_{101}$, and R$_{151}$, respectively.  Given the 4.14 km s$^{-1}$ pixel spacing of the APOGEE spectra, these CCF windows provide sensitivity to secondaries with velocity separations from the primary star of 106, 212, and 318 km s$^{-1}$.   

\par We used a combination of absolute and relative criteria to identify candidate SB2s based on the lowest R values they exhibited across all their APOGEE observations. Selecting candidates on the basis of their lowest observed R values allows us to identify systems even if they only exhibit a clear velocity separation in a single epoch of APOGEE spectra. Absolute criteria ensure that each star's CCF exhibits an asymmetry substantial enough to indicate the presence of a secondary star, while relative criteria based on ratios of the R values measured from different portions of the CCF (e.g., R$_{51}$, R$_{151}$, etc.) eliminate false positives due to sources whose CCFs exhibit significant asymmetries, but at velocities too large to be physically plausible for a bona fide SB2. 
We denote the smallest R value observed within a given CCF window across all a star's APOGEE observations as min$_{R_W}$ (where W indicates the width of the CCF window the R value was computed from, such that min$_{R_{151}}$ indicates the smallest R$_{151}$ observed for a given star).
To provide additional measures of the structure of each CCF, we also record the maximum response and bisector width of each CCF as $peak$ and $bisector_X$, respectively. Following the notation for the minimum R values across all epochs, we denote the maximum CCF response and bisector width across all observations as max$_{bisector_x}$ and max$_{peak}$, respectively.
Using these measures of the structure detected across all CCFs computed for a given source, we identify candidate SB2s with the following criteria:
\begin{itemize}
\item{To identify sources that exhibit a strong, central asymmetry on at least one epoch, we require: 
\begin{itemize}
\item{log$_{10}(\textrm{min}_{R_{101}}) <$ 0.83 \\ AND \\ 0.06 $<$ log$_{10} \frac{ \textrm{min}_{R_{151}}}{\textrm{min}_{R_{101}}} <$ 0.13}
\\ \\ OR
\item{log$_{10}(\textrm{min}_{R_{51}}) <$ 0.83 \\ AND \\ 0.05 $<$ log$_{10} \frac{ \textrm{min}_{R_{101}}}{\textrm{min}_{R_{51}}} <$ 0.2}
\end{itemize}}
\item{To eliminate sources with weak CCF responses, suggesting a poor template match, we require: 
\begin{itemize}
\item{log$_{10}(\textrm{max}_{peak}) > -0.5$ }
\end{itemize} }
\item{To eliminate sources whose CCF peaks are indicative of very low S/N or a reduction issue (i.e., too narrow or wide to be consistent with a single star or binary, or containing a greater degree of asymmetry than expected for 2-3 well detected CCF peaks), we require:
\begin{itemize}
\item{0.7 $>$ log$_{10}(\textrm{max}_{bisector_x}) > 2.3$ }
\item{log$_{10}(\textrm{min}_{R_{51}}) > 0.25$ }
\item{log$_{10}(\textrm{min}_{R_{101}}) > 0.22$ }
\end{itemize}}
\end{itemize}

These criteria identify 44 candidate M dwarf SB2s, or just more than 3\% of all 1350 M dwarf ancillary targets in the DR13 catalog. These targets are listed in Table \ref{table:SB2s}. Eight of these targets are among the 9 SB2s flagged by \citet{Deshpande2013} in their analysis of a subset of this sample, indicating that our methods are capable of recovering the majority of the short period, high flux ratio SB2s in the APOGEE database.  The exception is 2MJ19333940+3931372, for which the APOGEE CCFs show evidence for profile changes, but the secondary component does not cleanly separate from the primary peak in any of the three visits obtained by APOGEE. Modifying our selection criteria to capture this source as a candidate SB2 would significantly increase the number of false positives that would need to be removed from our sample via visual inspection, so we choose to retain our more conservative cuts that will produce a smaller, but higher fidelity, sample of candidate SB2s.
    
\subsection{Photometric Mass Estimates for Primary stars}    
    
We estimate the mass of 
the primary of each system in our sample using photometry and photometric calibrations from the literature.  Photometric mass estimates are valuable for multiple reasons: the presence of multiple components in the system's spectra renders the standard APOGEE/ASPCAP analysis unreliable, and the DR13 APOGEE parameters have been shown to be unreliable for even single M dwarfs \citep[see ][]{Souto2017}. For this photometric analysis, we adopted magnitudes from catalogs such as NOMAD \citep{Zacharias2005}, APOP \citep{Qi2015}, UCAC4 \citep{Zacharias2013}, UCAC5 \citep{Zacharias2017}, \citet{Viaux2013}, and \citet{Lepine2005}.  We did not attempt to infer or correct for stellar reddening in this process, as any extinction is expected to be minimal due to the stars' presence within the solar neighborhood. 



Stellar masses were derived using the ($V$-$K_s$) vs. mass color calibration derived by \citet{Delfosse2000}. 
For stars without a reliable $V$ magnitude reported in the literature, we adopted the $M_K$ absolute magnitude vs. mass calibration derived by \citet{Mann2015}. 
The absolute magnitudes were derived using distances in the literature. For the stars without distances reported we adopted $d$ = 20.0 pc.
The precision in the \citet{Delfosse2000} calibration is about 10\%, which returns an uncertainty of $\sim$ $\pm$ 0.05 $M(M_{\odot})$. 
    
\begin{deluxetable*}{lcrccccrr}
\tabletypesize{\tiny}
\tablecaption{Selected Binaries}
\tablewidth{0pt}
\tablehead{
\colhead{} & \colhead{Phot.} & \colhead{} & \colhead{} & \colhead{} & \colhead{} & \colhead{} & \colhead{} & \colhead{Well} \\
\colhead{} & \colhead{Mass} & \colhead{} & \colhead{} & \colhead{} & \colhead{} & \colhead{CCF} & \colhead{} & \colhead{Separated} \\
\colhead{2MASS ID} & \colhead{($M_{\odot}$)} & \colhead{Visits} & \colhead{$R_{151}$} & \colhead{$R_{101}$} & \colhead{$R_{51}$} & \colhead{maximum} & \colhead{$x_{range}$} & \colhead{Epochs} }
\startdata
2M00372323+4950469 & 0.207 & 3 & 6.09 & 7.84 & 5.88 & 0.32 & 78.60 & 0 \\
2M03122509+0021585 & 0.109 & 4 & 6.69 & 7.07 & 6.25 & 0.35 & 52.55 & 0 \\
2M03330508+5101297\tablenotemark{3} & 0.526 & 3 & 6.52 & 5.75 & 5.04 & 0.54 & 157.74 & 2 \\
2M03393700+4531160 & 0.268* & 6 & 3.75 & 3.28 & 2.88 & 0.59 & 31.36 & 4 \\
2M04281703+5521194\tablenotemark{1} & 0.168* & 13 & 7.60 & 6.75 & 4.96 & 0.42 & 186.37 & 0 \\
2M04373881+4650216 & 0.438* & 4 & 8.60 & 7.20 & 5.19 & 0.66 & 14.90 & 2 \\
2M04595013+3638144 & 0.203 & 3 & 5.78 & 5.40 & 3.96 & 0.39 & 15.48 & 3 \\
2M05421216+2224407 & 0.178 & 4 & 5.18 & 4.67 & 3.29 & 0.44 & 14.10 & 1 \\
2M05504191+3525569 & 0.153* & 3 & 6.74 & 5.85 & 4.42 & 0.52 & 48.04 & 1 \\
2M06115599+3325505\tablenotemark{1} & 0.152 & 13 & 4.29 & 3.47 & 3.08 & 0.69 & 27.22 & 11 \\
2M06125378+2343533 & 0.562* & 3 & 9.38 & 8.01 & 5.80 & 0.77 & 32.13 & 3 \\
2M06213904+3231006 & 0.430 & 6 & 4.19 & 3.54 & 2.67 & 0.68 & 26.13 & 5 \\
2M06561894-0835461 & 0.193 & 4 & 6.75 & 6.22 & 4.90 & 0.61 & 31.88 & 3 \\
2M07063543+0219287\tablenotemark{2} & 0.653* & 3 & 5.48 & 4.46 & 3.18 & 0.83 & 7.38 & 1 \\
2M07444028+7946423 & 0.601* & 3 & 3.17 & 2.58 & 3.22 & 0.76 & 13.50 & 2 \\
2M08100405+3220142 & 0.376 & 6 & 5.90 & 4.91 & 3.42 & 0.79 & 19.03 & 3 \\
2M08351992+1408333\tablenotemark{3} & 0.149 & 3 & 8.25 & 7.01 & 5.13 & 0.47 & 12.03 & 1 \\
2M10331367+3409120\tablenotemark{3} & 0.515 & 3 & 4.25 & 3.66 & 2.93 & 0.77 & 14.51 & 2 \\
2M10423925+1944404 & 0.403 & 4 & 6.76 & 5.70 & 4.10 & 0.50 & 12.47 & 3 \\
2M10464238+1626144\tablenotemark{1} & 0.181 & 3 & 5.00 & 4.83 & 3.68 & 0.46 & 14.51 & 3 \\
2M10520326+0032383\tablenotemark{4} & 0.175 & 3 & 2.46 & 2.06 & 3.16 & 0.43 & 8.73 & 3 \\
2M11081979+4751217 & 0.191 & 5 & 4.66 & 4.14 & 2.99 & 0.47 & 107.92 & 5 \\
2M12045611+1728119\tablenotemark{1} & 0.386 & 3 & 6.40 & 5.70 & 4.43 & 0.61 & 25.47 & 3 \\
2M12193796+2634445 & 0.266 & 8 & 8.76 & 7.97 & 5.98 & 0.38 & 72.52 & 0 \\
2M12214070+2707510 & 0.465 & 11 & 5.93 & 4.86 & 3.38 & 0.78 & 32.00 & 6 \\
2M12260547+2644385 & 0.926* & 11 & 8.82 & 7.24 & 5.19 & 0.89 & 5.27 & 0 \\
2M12260848+2439315 & 0.348 & 8 & 4.33 & 3.76 & 2.72 & 0.60 & 53.74 & 7 \\
2M14545496+4108480 & 0.202 & 4 & 5.05 & 4.39 & 3.03 & 0.48 & 36.58 & 4 \\
2M14551346+4128494 & 0.340 & 4 & 7.45 & 6.28 & 5.30 & 0.70 & 13.67 & 4 \\
2M14562809+1648342 & 0.542 & 3 & 9.85 & 8.49 & 6.22 & 0.79 & 11.20 & 1 \\
2M15183842-0008235\tablenotemark{3} & 0.528* & 3 & 5.49 & 4.44 & 3.20 & 0.83 & 10.74 & 1 \\
2M15192613+0153284\tablenotemark{1} & 0.221 & 14 & 9.48 & 8.54 & 6.37 & 0.52 & 22.29 & 0 \\
2M15225888+3644292\tablenotemark{3}\tablenotemark{5} & 0.644* & 3 & 6.73 & 5.54 & 3.90 & 0.86 & 5.69 & 1 \\
2M17204248+4205070\tablenotemark{1} & 0.158 & 15 & 7.02 & 6.73 & 5.21 & 0.56 & 24.15 & 12 \\
2M18514864+1415069 & 0.479* & 3 & 10.64 & 8.95 & 6.70 & 0.76 & 30.77 & 0 \\
2M19081153+2839105 & 0.184 & 13 & 5.55 & 4.79 & 3.41 & 0.60 & 49.25 & 1 \\
2M19235494+3834587\tablenotemark{1}\tablenotemark{2} & 0.822* & 3 & 3.19 & 2.56 & 2.12 & 0.71 & 34.93 & 1 \\
2M19433790+3225124 & 0.630* & 3 & 8.10 & 6.93 & 5.21 & 0.86 & 28.96 & 1 \\
2M19560585+2205242 & 0.168 & 14 & 9.98 & 9.14 & 6.70 & 0.60 & 14.57 & 0 \\
2M20474087+3343054\tablenotemark{2} & 0.631* & 3 & 4.87 & 4.11 & 3.16 & 0.83 & 17.71 & 2 \\
2M21005978+5103147 & 0.380 & 5 & 5.35 & 4.44 & 3.03 & 0.69 & 8.47 & 4 \\
2M21234344+4419277 & 0.494 & 8 & 4.56 & 3.62 & 4.03 & 0.59 & 21.84 & 7 \\
2M21442066+4211363\tablenotemark{1} & 0.149* & 12 & 2.62 & 4.36 & 3.21 & 0.47 & 62.68 & 12 \\
2M21451241+4225454 & 0.212 & 12 & 8.05 & 7.38 & 5.27 & 0.64 & 11.33 & 0 \\
\enddata
\tablecomments{Stellar masses are estimated from the (V-K) vs. Mass relation derived by \citet{Delfosse2000}; values tagged with a * are determined from the M$_{K}$ vs. Mass relation derived by \citet{Mann2015}, after adopting a distance based on a measured trigonometric parallax or a fiducial solar neighborhood distance of 20 pc.}
\tablenotetext{1}{Identified by \citet{Deshpande2013} as an SB2.}
\tablenotetext{2}{Identified by \citet{ElBadry2018} as an SB2.}
\tablenotetext{3}{Found here to be an SB3.}
\tablenotetext{4}{Found here to be an SB4.}
\tablenotetext{5}{Identified by \citet{ElBadry2018} as an SB3.}
\tablecaption{}
\label{table:SB2s}
\end{deluxetable*}
    
    \subsection{\textit{Additional RV monitoring with HET/HRS\label{hethrs}}}
   
    We supplemented the APOGEE observations for a few systems with visible light spectroscopy from the fiber-fed High Resolution Spectrograph (hereafter, HRS; \citet{Tull1998}) on the 9.2 meter Hobby-Eberly Telescope (hereafter, HET; \citet{Ramsey1998}). We used HRS with the 316g5936 cross-disperser in the 30K resolution mode with the 2 arcsecond slit and the central grating angle. This produced spectra spanning the wavelength range from 4076 -- 7838 nm, although we only used the region from $\sim6600$ nm redward because these M dwarf spectra suffer from low signal-to-noise at shorter wavelengths. All observations were conducted in queue mode \citep{Shetrone2007}. We exposed for 10-20 minutes per target, per epoch, based on magnitude. Wavelength calibration was obtained from ThAr frames that bracket the observation.
    
    Spectra were extracted using a custom optimal extraction pipeline, modeled after the SpeXTool pipeline developed by \citet{Cushing2004} and similarly written in the Interactive Data Language (IDL). The HRS pipeline automates basic image processing procedures, such as overscan correction, bias subtraction, flat-fielding, and core spectral extraction processes such as tracing each order, computing the optimal fiber profile, and extracting source and ThAr lamp spectra.  Wavelength solutions are derived by fitting a multi-order function to the ThAr spectra using the linelist reported by \citep{Murphy2007} and applied to the object spectra.  
    
    Extracted, wavelength calibrated spectra were then merged across areas of inter-order interlap, and trimmed to exclude regions of significant contamination by telluric absorption or OH night-sky emission lines.  Regions dominated by telluric absorption were identified by inspecting the LBLRTM atmospheric model \citep{Clough2005}; sharp night-sky emission features were removed by linearly interpolating over wavelength regions known to host strong emission lines \citep[e.g.][]{Abrams1994}.  
    
\section{BULK ANALYSIS}

	\subsubsection{\textit{Cuts}}
	Of the 44 sources that we identified as likely SB2s, 9 systems 
do not exhibit, at any epoch for which we have data, a velocity separation sufficiently large to reliably measure the RVs of both components with our initial RV extraction method. 
Analysis with TODCOR allowed us to recover RVs for 1 of these 9 systems, providing a sample of 36 multiples with RVs for futher analysis. Seven of these 36 systems are higher order systems (6 triples and 1 quadruple) with moderate velocity separations but poorly determined RVs due to significant blending in one or more of the APOGEE observations. TODCOR analysis allowed us to recover RVs for 5 of the 6 triples; the quadruple system remains unsolved. Exclusion of the 8 poorly separated systems (see Table 2), and the two unsolved higher order multiples\footnote{2M10331367+3409120, 2M10520326+0032383} leave 34 targets for which we are able to measure mass ratios. 

  \begin{deluxetable}{cc}
  \tabletypesize{\scriptsize}
  \tablecaption{Excluded Targets}
  \tablewidth{0pt}
  \tablehead{\colhead{2MASS ID} & \colhead{Max $\Delta$RV ($\frac{km}{s}$)}}
  \startdata
  2M00372323+4950469 & 22.58 \\
  2M03122509+0021585 & 15.13 \\
  2M04281703+5521194 & 30.77 \\
  2M12193796+2634445 & 30.05 \\
  2M15192613+0153284 & 28.48 \\
  2M18514864+1415069 & 24.97 \\
  2M19560585+2205242 & 40.27 \\
  2M21451241+4225454 & 15.04 \\
  \enddata
  \tablecaption{}
  \label{table:exclusions}
  \end{deluxetable}

	\subsection{\textit{RV extraction from APOGEE visits}}
    Radial velocities were extracted from APOGEE CCFs for all components of each system using the procedures developed by \citet{Fernandez2017}. We describe the process briefly here, but refer the reader to the earlier work for a detailed description. Radial velocities were extracted from each APOGEE visit CCF using a multi-step fitting process, after converting the CCF's abscissa from lag space to velocity space.
    \par In the first step, a Lorentzian was fit to the maximum peak of the CCF. This Lorentzian was then subtracted from the CCF, removing the primary peak. With the primary peak removed, a second Lorentzian was then fit to the maximum in the residual CCF, which was implicitly identified as the secondary peak. For sources with multiple APOGEE visit spectra, the epoch containing the greatest separation between the primary and secondary peaks was identified as the ``widest separated CCF''. A dual-Lorentzian model was then fit to the widest separated CCF using the peak centers identified earlier for the primary and secondary components to initialize the fit. Finally the dual-Lorentzian fit was performed on the remaining epochs using the peak heights and widths measured from the ``widest separated epoch'' to initialize the fit, along with the previously identified peak velocities.
    \par A notable deficiency of this extraction method is that the resultant RVs lack an individually defined uncertainty value. Section 3.3 of F17 details their calculation of a pseudo-normal 1$\sigma$ error of $\sim$1.8$\frac{km}{s}$. We adopt this ensemble uncertainty value for all RVs extracted by CCF-fitting. Another difficulty the CCF fitting method faces is consistent assignment of velocities to the primary and secondary components for SB2s with flux ratios close to unity. The accuracy of the RV values measured via this extraction technique suffered for epochs with small velocity separations, so we flagged these systems for follow up analysis with the TODCOR algorithm \citep{Zucker1994}, which is more adept at extracting velocities from epochs with small velocity separations. The CCF-fit derived RVs are replaced at any epochs for which TODCOR RVs were extracted.
	\par
\begin{figure*}
\centerline{\includegraphics[scale=.60]{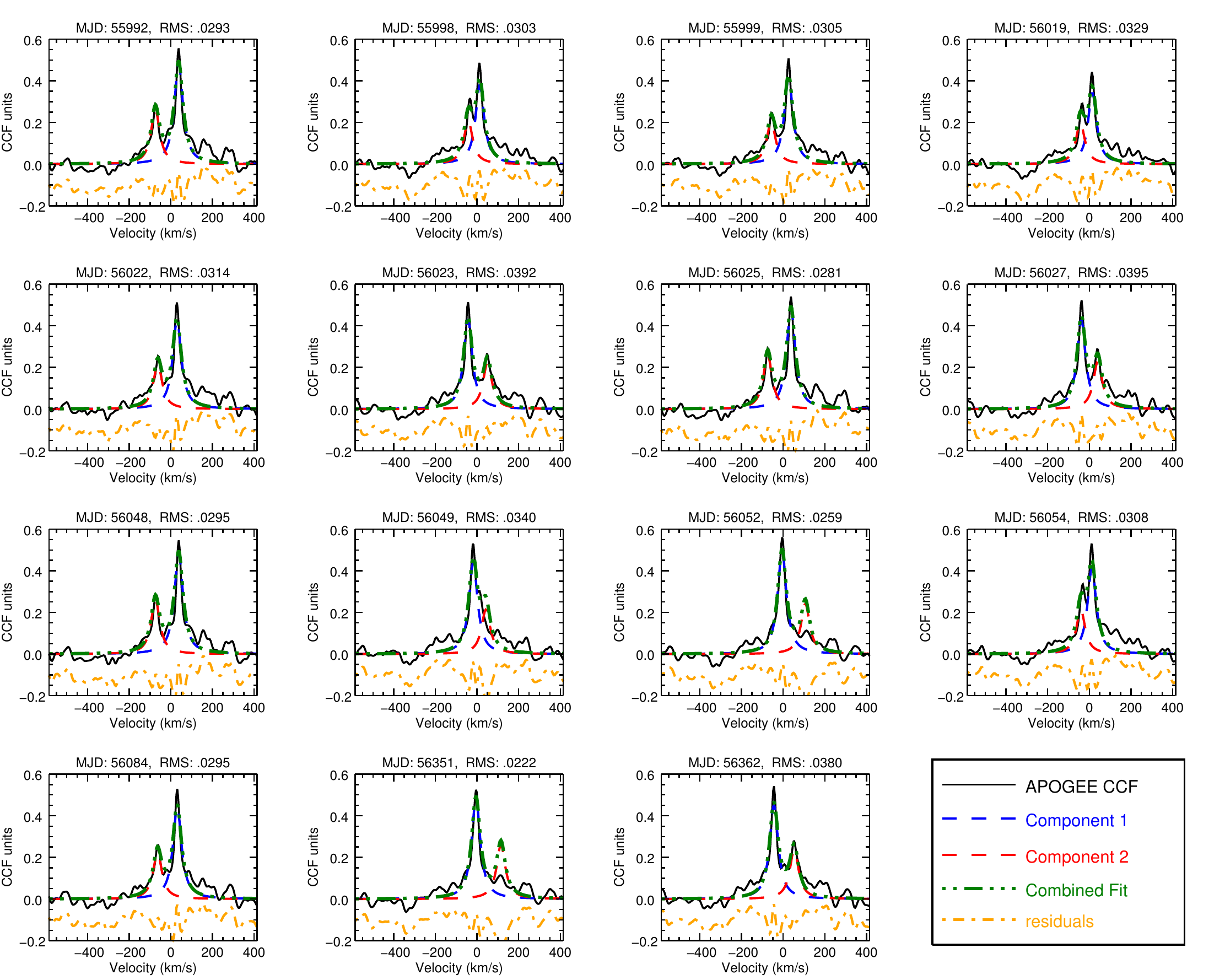}}
\caption{APOGEE CCFs for 2M17204248+4205070, transformed from lag units into velocity space. The blue and red dashed lines represent the Lorentzian fits to the primary and secondary velocity peaks, respectively; the sum of the two fits is shown as a green dashed-dotted line, and the residuals of the fit are shown as an orange dashed line, offset by 0.1 for clarity. Note the spurious secondary fits in the third and fourth rows, at epochs with MJDs of 56049, 56052 and 56351. Erroneous velocity values associated with spurious fits such as these have been removed from our dataset by visual inspection.}
\label{fig:1720+4205_CCFs}
\end{figure*}

 Figure \ref{fig:1720+4205_CCFs} shows the Lorentzian fits to the primary and secondary peaks in all CCFs computed from APOGEE spectra of 2M17204248+4205070. Figures such as this were visually inspected to identify cases where the fits to the CCF peaks were obviously incorrect (i.e., a fit to a spurious structure in the CCF, most often occurring at epochs without well separated CCF peaks). In such cases, spurious RV measures were removed from the sample. SB2 radial velocities are listed in Table 3, which is presented here as a stub. The full version can be found in Appendix B. 
    
\startlongtable
\begin{deluxetable*}{crccrrr}
\tabletypesize{\tiny}
\tablecaption{Radial Velocity Measurements of SB2s}
\tablewidth{0pt}
\tablehead{
\colhead{2MASS ID} & \colhead{Visit} & \colhead{Epoch (MJD)} & \colhead{SDSS plate \& Fiber} & \colhead{SNR} & \colhead{v$_{prim}(\frac{km}{s})$} & \colhead{v$_{sec}(\frac{km}{s})$}}
\startdata
2M03393700+4531160 & 1 & 56195.3409 & 6244-56195-086 & 117 & -16.2 & 33.8 \\
$\vert$ & 2 & 56200.2983 & 6244-56200-131 & 210 & -20.4 & 38.9 \\
$\vert$ & 3 & 56223.2868 & 6244-56223-131 & 215 & -37.3 & 53.8 \\
$\vert$ & 4 & 56196.3190 & 6245-56196-077 & 168 & 52.0 & -36.3 \\
$\vert$ & 5 & 56202.2755 & 6245-56202-074 & 137 & 7.5 & - \\
$\vert$ & 6 & 56224.3188 & 6245-56224-077 & 186 & 11.1 & 3.9 \\
2M04373881+4650216 & 1 & 56176.4835 & 6212-56176-050 & 49 & -40.8 & -44.4 \\
$\vert$ & 2 & 56234.3042 & 6212-56234-050 & 32 & -31.5 & -56.4 \\
$\vert$ & 3 & 56254.2442 & 6212-56254-050 & 62 & -26.1 & -63.8 \\
$\vert$ & 4 & 56260.2176 & 6212-56260-050 & 44 & -26.6 & -61.6 \\
: & : & : & : & : & : & : \\
\enddata
\tablecomments{Dashed out velocities indicate spurious RVs omitted from analysis. RVs not extracted via TODCor are assigned the ensemble uncertainty of $\sim 1.8 \frac{km}{s}$.}
\tablecaption{}
\label{table:SB2_RVs}
\end{deluxetable*}

\subsection{RV extraction via TODCOR}
    
We used the TODCOR algorithm \citep{Zucker1994} to measure RVs from all HET/HRS spectra and any APOGEE spectra flagged with low RV separations. This TODCOR analysis followed the procedures previously discussed by \citet{Bender2005} and used the algorithm implementation of \citet{Bender2012}; we briefly summarize here the key parts of this implementation and its modification for use with APOGEE spectra, but refer the reader to the previous presentations for more details. TODCOR simultaneously cross-correlates each target spectrum against the spectra of two template stars.  For both the HRS and APOGEE observations we generated template spectra from the BT-Settl library \citep{Allard2012}, convolved to each spectrograph's resolution and rotationally broadened using the \citet{Claret2000} non-linear limb darkening models.  Templates were optimized for each binary by maximizing the peak correlation, using a template grid with $\Delta T_{eff}=100 K$, $\Delta \log(g)=0.5$, and $\Delta [M/H]=0.5$. This optimization happens independently for the HRS and APOGEE spectra.  Due to variations in the quality of the linelists that underlie the BT-Settl models, we frequently derive slightly different optimal template sets for visible and near-infrared spectra.  These differences are typically within one or two gridpoints (i.e., 100-200 K in temperature, and $<$0.5 dex in $\log(g)$ and [M/H]), and RVs are not sensitive to template choice at this resolution.  An initial set of RVs was derived for all epochs of a given system by using the secondary-to-primary flux ratio ($\alpha$) optimization of TODCOR.  These RVs were then iterated with a fixed $\alpha$ corresponding to the average $\alpha$ derived for all epochs.  We extract component RVs from a quadratic fit to the top six-to-eight points of the 2D cross-correlation peak, and derive uncertainties using the maximum likelihood formalism of \citet{Zucker1994}. We flag the epochs for which velocities were extracted from HET/HRS or APOGEE spectra using this technique.

Our analysis also revealed several triple systems composed of a short period binary along with a wider companion.  We analyzed these systems with the TRICOR extension of TODCOR \citep{Zucker1995}, using three templates and solving for two independent flux ratios.  The small number of epochs and relatively small temporal baselines limit what we can conclude about the stability of these systems or the orbital period of the wide component.  We present the RVs derived for these candidate SB3 systems in Table \ref{table:SB3_RVs}; in all cases, the RVs derived for the wide companion are consistent within the uncertainties with the systemic RV of the inner pair, as would be expected for bound systems.  Additionally, we did not detect any significant RV motion from any of the apparent wide companions.  Consequently, we conclude that these are likely to be bound, hierarchical triple systems.

\begin{deluxetable*}{crccrrcrcrc}
\tabletypesize{\scriptsize}
\tablecaption{Radial Velocity Measurements of SB3s}
\tablewidth{0pt}
\tablehead{\colhead{2MASS ID} & \colhead{Visit} & \colhead{Epoch (MJD)} & \colhead{SDSS plate \& Fiber} & \colhead{SNR} & \colhead{v$_{prim}(\frac{km}{s})$} & \colhead{$\sigma_{prim}$} & \colhead{v$_{sec}(\frac{km}{s})$} & \colhead{$\sigma_{sec}$} & \colhead{v$_{ter}(\frac{km}{s})$} & \colhead{$\sigma_{ter}$}}
\startdata
2M03330508+5101297 & 1 & 56257.1757 & 6538-56257-087 & 42 & 30.86 & 0.53 & -84.72 & 1.57 & -11.49 & 1.02 \\
$\vert$ & 2 & 56261.1912 & 6538-56261-088 & 73 & -65.39 & 0.30 & 78.53 & 0.99 & -10.98 & 0.73 \\
$\vert$ & 3 & 56288.1033 & 6538-56288-069 & 39 & -14.83 & 0.47 & -4.44 & 1.10 & -10.48 & 0.93 \\
2M04595013+3638144 & 1 & 56256.2342 & 6542-56256-294 & 43 & -29.49 & 0.39 & 33.45 & 1.52 & -10.06 & 0.53 \\
$\vert$ & 2 & 56262.2454 & 6542-56262-187 & 44 & -33.77 & 0.46 & 39.26 & 1.43 & -10.74 & 0.59 \\
$\vert$ & 3 & 56288.1841 & 6542-56288-192 & 52 & -33.04 & 0.44 & 38.62 & 1.47 & -8.80 & 0.71 \\
2M08351992+1408333 & 1 & 56284.3821 & 6612-56284-106 & 248 & 37.04 & 0.36 & -17.11 & 0.84 & 17.30 & 0.48 \\
$\vert$ & 2 & 56290.5157 & 6612-56290-105 & 246 & 5.62 & 0.42 & 27.05 & 1.19 & 18.66 & 0.53 \\
$\vert$ & 3 & 56315.3140 & 6612-56315-105 & 234 & -0.02 & 0.39 & 33.29 & 0.79 & 20.10 & 0.42 \\
2M15183842-0008235 & 1 & 56080.2737 & 5906-56080-244 & 235 & 6.30 & 0.27 & -86.25 & 0.63 & -36.58 & 0.30 \\
$\vert$ & 2 & 56435.2488 & 5906-56435-244 & 241 & -46.42 & 0.19 & -18.39 & 0.42 & -37.26 & 0.24 \\
$\vert$ & 3 & 56467.1400 & 5906-56467-256 & 248 & -41.37 & 0.28 & -23.34 & 0.50 & -35.66 & 0.26 \\
2M15225888+3644292 & 1 & 56735.3805 & 5756-56735-106 & 152 & -68.10 & 0.43 & -19.08 & 2.10 & -56.64 & 0.74 \\
$\vert$ & 2 & 56740.4314 & 5756-56740-009 & 185 & -82.70 & 0.24 & 8.13 & 2.50 & -57.52 & 0.50 \\
$\vert$ & 3 & 56762.3135 & 5756-56762-010 & 157 & -15.30 & 0.30 & -136.30 & 1.70 & -55.46 & 0.60 \\
\enddata
\tablecomments{All velocities in this table were extracted via TODCOR.}
\tablecaption{}
\label{table:SB3_RVs}
\end{deluxetable*}

\subsection{\textit{Mass Ratio Measurements}} 
    \subsubsection{Measurement Technique}
\begin{figure*}[t!]
\centerline{\includegraphics[scale=0.5]{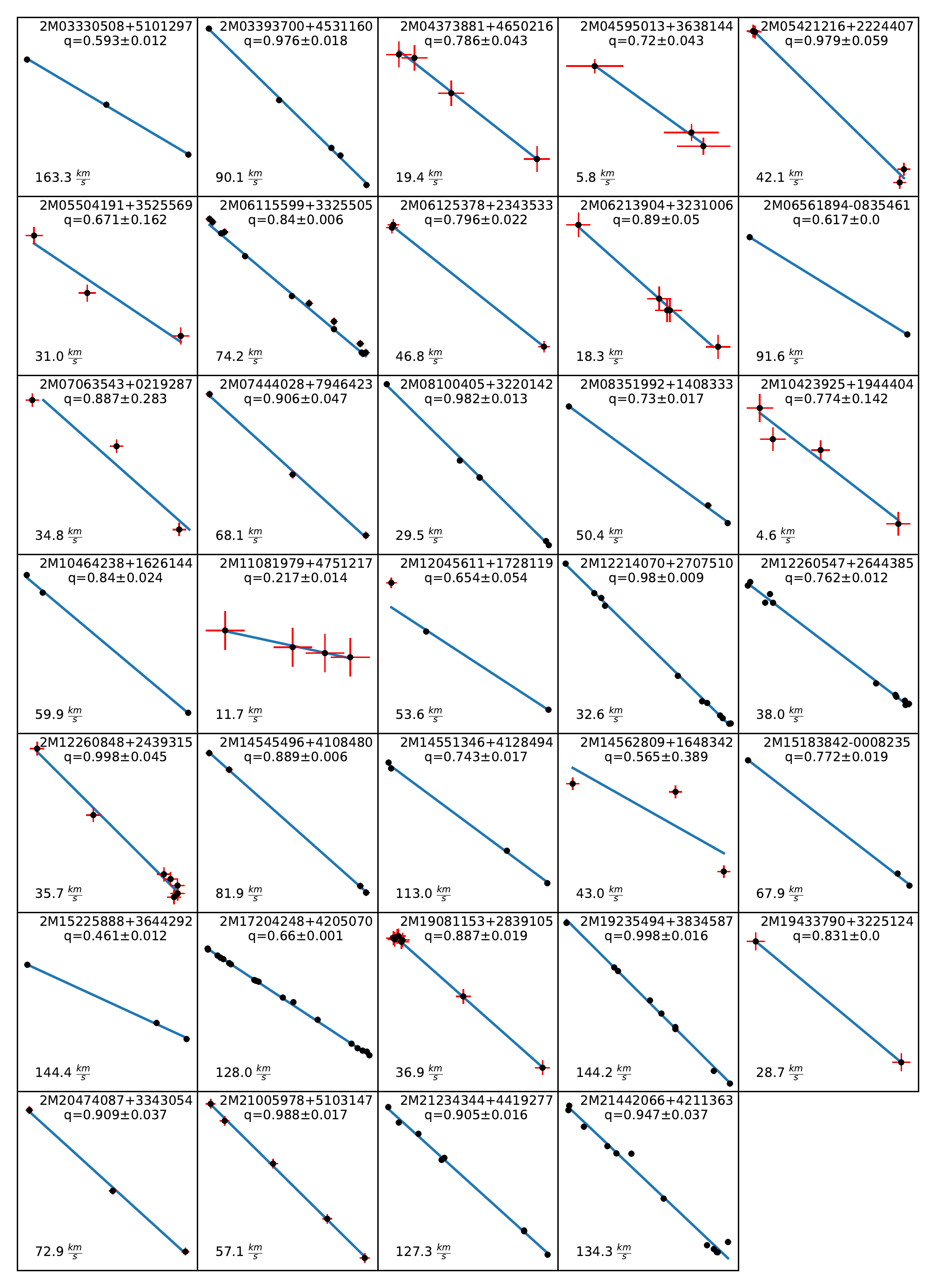}}
\caption{\small{Wilson plots of each mass ratio measurement in our sample. The black circles are radial velocity observations. Uncertainties are shown as red bars. The blue line of each plot is best fit line to the data, from which the mass ratio is calculated. The horizontal axis is $v_{sec}$, the vertical axis is $v_{prim}$. The aspect ratio between the vertical and horizontal axes of each subplot is 1. In the lower left corner of each subplot the $v_{sec}$ range is given.}}
\label{fig:WilsonGrid}
\end{figure*}

\begin{figure}[t!] 
\centerline{\includegraphics[scale=0.45]{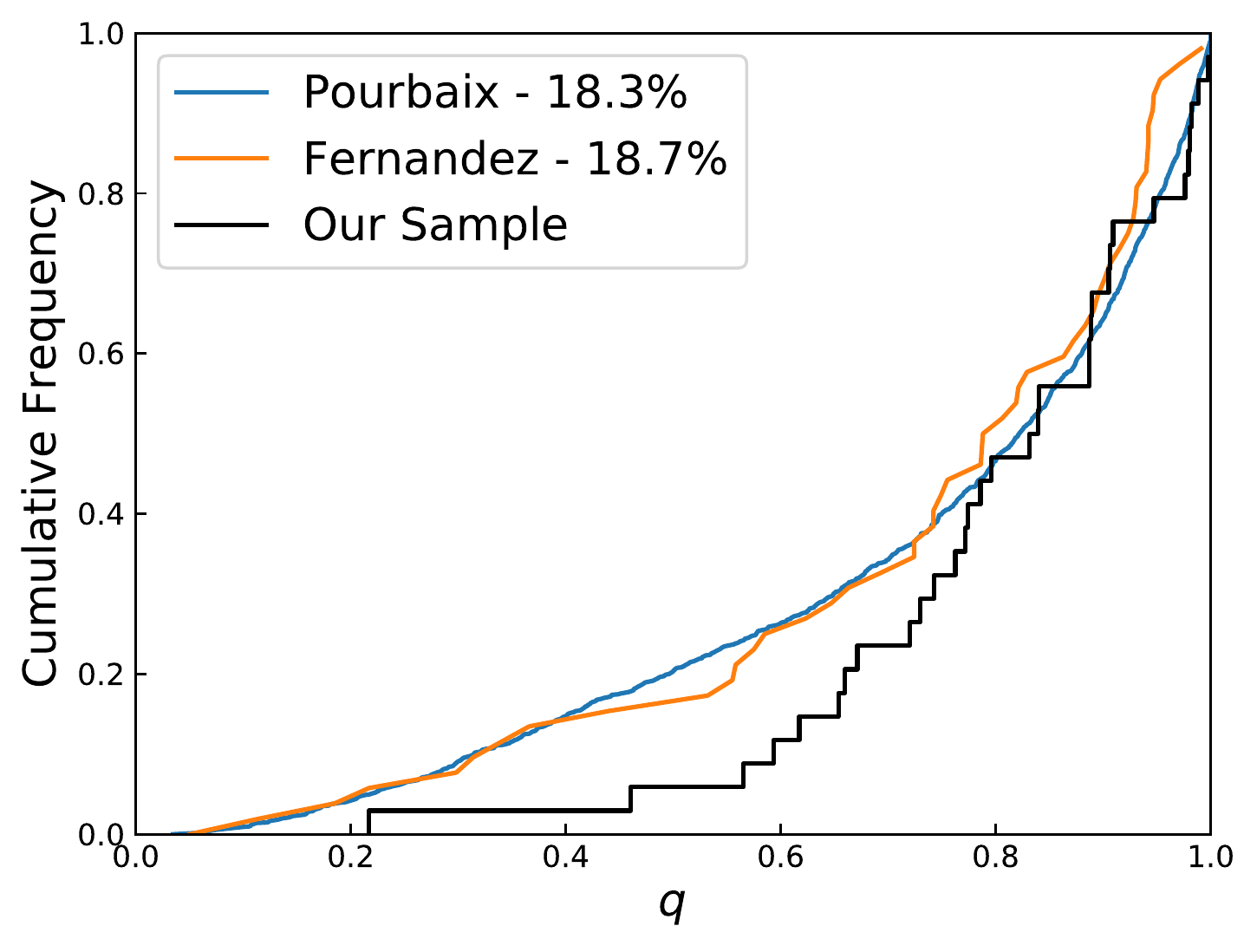}}
\caption{\small{Mass ratio distribution of our sample, and the Kolmogorov-Smirnov statistics comparing our distribution and those of \citet{Pourbaix2004} and \citet{Fernandez2017}}}
\label{fig:MassRatioDistribution}
\end{figure}

    We used the method presented in \citet{Wilson1941} to measure the mass ratios ($q \equiv\frac{M_{sec}}{M_{prim}}$) of our targets via a linear regression of primary velocity as a function of secondary velocity. 
Mass ratios greater than 1 were measured for 4 targets in the sample of 34 with reliably extracted RVs; we interpret these high q values as a sign of a primary/secondary assignment mismatch, and have therefore swapped the assignment for these targets, and included the resulting lower q values in the remainder of our analysis. 
The mass ratios measured for the 34 member sample are listed in Table \ref{table:MassRatios}, with Wilson plots for each system presented in Figure \ref{fig:WilsonGrid}, and the mass ratio distribution of the entire sample shown in Figure \ref{fig:MassRatioDistribution}. 

\begin{deluxetable}{llcr}
\tabletypesize{\tiny}
\tablecaption{Mass ratio and $\Delta$RV of analyzed stars}
\tablewidth{0pt}
\tablehead{\colhead{2MASS ID} & \colhead{$q\pm\delta q$} & \colhead{$\frac{\delta q}{q}$} & \colhead{Max $\Delta$RV}}
\startdata
2M03330508+5101297 & $0.593\pm0.012$ & 0.020 & 143.92 \\
2M03393700+4531160 & $0.976\pm0.018$ & 0.018 & 91.13 \\
2M04373881+4650216 & $0.786\pm0.043$ & 0.054 & 37.69 \\
2M04595013+3638144 & $0.720\pm0.043$ & 0.060 & 73.03 \\
2M05421216+2224407 & $0.979\pm0.059$ & 0.060 & 41.65 \\
2M05504191+3525569 & $0.671\pm0.162$ & 0.241 & 37.27 \\
2M06115599+3325505 & $0.840\pm0.006$ & 0.008 & 70.40 \\
2M06125378+2343533 & $0.796\pm0.022$ & 0.028 & 42.88 \\
2M06213904+3231006\tablenotemark{1} & $0.890\pm0.050$ & 0.044 & 86.97 \\
2M06561894-0835461\tablenotemark{2} & $0.617\pm-$ & - & 82.85 \\
2M07063543+0219287\tablenotemark{1} & $0.887\pm0.283$ & 0.251 & 57.98 \\
2M07444028+7946423\tablenotemark{1} & $0.906\pm0.047$ & 0.042 & 91.12 \\
2M08100405+3220142 & $0.982\pm0.013$ & 0.013 & 34.01 \\
2M08351992+1408333 & $0.730\pm0.017$ & 0.024 & 54.15 \\
2M10423925+1944404 & $0.774\pm0.142$ & 0.184 & 29.56 \\
2M10464238+1626144 & $0.840\pm0.024$ & 0.028 & 68.10 \\
2M11081979+4751217 & $0.217\pm0.014$ & 0.066 & 64.32 \\
2M12045611+1728119 & $0.654\pm0.054$ & 0.082 & 141.45 \\
2M12214070+2707510\tablenotemark{1} & $0.980\pm0.009$ & 0.009 & 42.65 \\
2M12260547+2644385 & $0.762\pm0.012$ & 0.016 & 40.54 \\
2M12260848+2439315 & $0.998\pm0.045$ & 0.045 & 41.42 \\
2M14545496+4108480 & $0.889\pm0.006$ & 0.007 & 99.39 \\
2M14551346+4128494 & $0.743\pm0.017$ & 0.023 & 99.42 \\
2M14562809+1648342 & $0.565\pm0.389$ & 0.688 & 53.92 \\
2M15183842-0008235 & $0.772\pm0.019$ & 0.025 & 92.55 \\
2M15225888+3644292 & $0.461\pm0.012$ & 0.025 & 121.00 \\
2M17204248+4205070 & $0.660\pm0.001$ & 0.002 & 110.28 \\
2M19081153+2839105 & $0.887\pm0.019$ & 0.022 & 47.71 \\
2M19235494+3834587 & $0.998\pm0.016$ & 0.016 & 161.26 \\
2M19433790+3225124\tablenotemark{2} & $0.831\pm-$ & - & 58.67 \\
2M20474087+3343054 & $0.909\pm0.037$ & 0.041 & 76.92 \\
2M21005978+5103147 & $0.988\pm0.017$ & 0.017 & 57.22 \\
2M21234344+4419277 & $0.905\pm0.016$ & 0.018 & 124.27 \\
2M21442066+4211363 & $0.947\pm0.037$ & 0.039 & 124.23 \\
\enddata
\tablenotetext{1}{For these targets $q > 1$. We assume this is due to a primary/secondary mismatch, and report $q^{-1}$ as $q$}
\tablenotetext{2}{Only two epochs were usable for these targets, therefore $\delta q$ is not well defined.}
\tablecaption{}
\label{table:MassRatios}
\end{deluxetable}
    
    
    \subsection{\textit{Detection limits}}\label{sec:detection_limits}
    \par 
A detailed analysis of the selection effects introduced by the cadence of APOGEE observations and the sensitivity of our binary detection method to systems with varying mass ratios, inclinations, and separations is beyond the scope of this paper. To provide first order indicators of the biases and limits that affect the make-up of our sample, however, we calculate fiducial detection limits imposed by the requirement that we detect multiple components, well-separated in velocity space, in at least one APOGEE spectrum. It is worth noting that the ancillary science program is composed of targets of opportunity, and was not designed to be complete in distance or magnitude.

\begin{figure}[t!] 
\centerline{\includegraphics[scale=0.45]{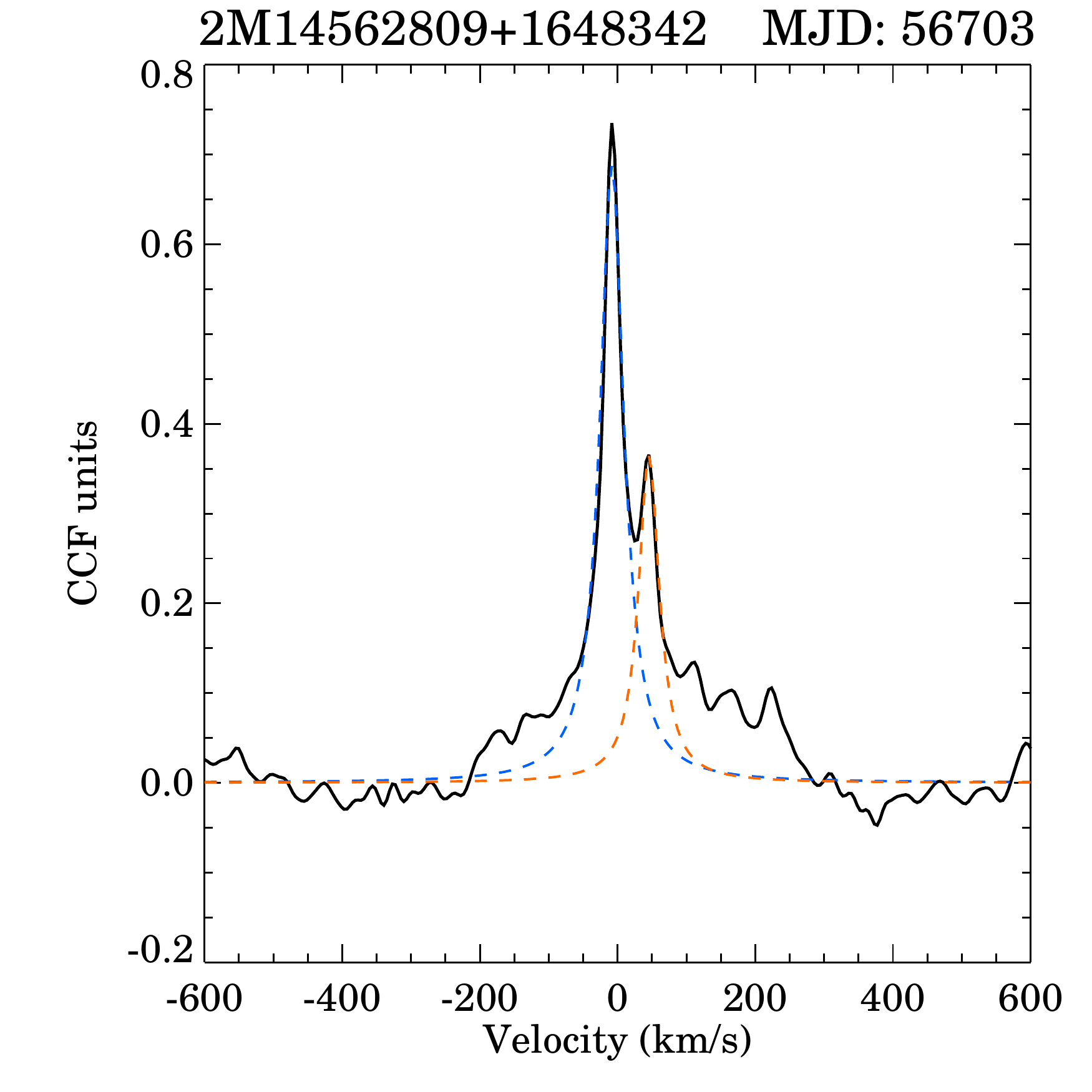}}
\caption{\small{An APOGEE CCF for 2M14562809+1648342, a binary whose mass ratio ($q=0.565$) places it near the fiducial $q=0.5$ / $\Delta$H$=1.2$ limit that we estimate for where our search method will become substantially incomplete due solely to the inability to confidently detect the secondary companion, even at high velocity separations.}}
\label{fig:fiducialCCFdetectionlimit}
\end{figure}

We first consider the H-band flux ratio that a system must satisfy to be detectable as a double-lined spectroscopic binary. Given the characteristic scale of the substructure in a typical APOGEE CCF, and the typical ratios of the integrated areas of the primary \& secondary CCF peaks for systems identified by our detection routine, we estimate that our detection method will begin to become significantly incomplete for systems with a secondary-to-primary H-band flux ratio $\leq$ 0.2, or a magnitude difference of $\Delta H \leq$ 1.2. Consulting the H-band magnitudes and fluxes tabulated by \citet{Kraus2007}, we find that binaries with M dwarf primaries typically satisfy this H-band flux ratio limit if they possess a mass ratio $\geq$ 0.5. To illustrate the nature of this detection limit, we show in Figure \ref{fig:fiducialCCFdetectionlimit} the CCF of a system with a mass ratio near, but just above, this fiducial $q=0.5$ limit.

We next consider the limits on system separation and orbital period imposed by the requirement that we detect two clearly separated CCF peaks. 
Given the resolution of the APOGEE spectrograph, a system's CCF peaks are clearly separated for primary-secondary velocity separations of $\Delta RV_{lim}\sim$30 km s$^{-1}$ or more \citep{Fernandez2017}. This velocity separation threshold imposes a joint constraint on the inclination ($i$), total mass (M$_1$ + M$_2$) and orbital period ($P$) of systems in our sample:

    $$ P \leq \frac{2~\pi~G~(M_1 + M_2)~\textrm{sin}^3~i}{\Delta RV_{lim}^3} $$

Expressing this constraint in terms of the system's semi-major axis ($a$), rather than its orbital period, makes this limit:

    $$ a \leq \frac{G~(M_1 + M_2)~\textrm{sin}^3 i}{\Delta RV_{lim}^2} $$

Edge-on ($i\sim90^{\circ}$), equal-mass systems are the most favorable configuration for detection: for a fiducial pair of 0.5 M$_{\odot}$ stars, we find a limiting period of $\sim$1 year and a limiting separation of 1 AU; for a lower mass pair of 0.25 M$_{\odot}$ stars, we find a limiting period and separation of $\sim$0.5 years \& AU, respectively\footnote{As years and AUs are defined based on the properties of our own solar system, and scale identically with system mass and inclination, the limiting period and separation for a fiducial system will be numerically identical when expressed in units of years and AUs}. Due to the sin$^3 i$ term in these limits, however, these limits decrease quickly with inclination: a modest inclination of 30$^{\circ}$ reduces the detection limits for the 0.5 M$_{\odot}$ binary to $\sim$0.12 years and AU, and to 0.06 years and AU for the 0.25 M$_{\odot}$ system.

The considerations above demonstrate that our sample is biased towards edge-on systems with mass ratios $\geq$0.5, and will be most complete for systems with characteristic periods and separations of $\leq$0.1 years and $\leq$0.1 AUs. We therefore adopt 0.1 years and 0.1 AUs as useful benchmarks for the completeness limits of our observed sample, and for comparing the properties of this sample to those measured from other samples of binary stars reported in the literature.

\section{FULL ORBIT FITS FOR HIGH VISIT, HIGH $\Delta$RV SYSTEMS}
    \subsection{\textit{Criteria for full orbital fits}}

The choice of targets chosen for orbital fits was made using the 3 following criteria, met by 4 systems:

\begin{itemize}

\item{Primary and secondary RVs for at least 8 visits.}

\item{Fractional mass ratio uncertainty less than $10\%$.}

\item{$V_{cov}$ value of at least 0.7, where $V_{cov}$ is the velocity coverage statistic presented in equation 5 of \citet{Fernandez2017}.
$$V_{cov}=\frac{N}{N-1}\Bigg(1-\frac{1}{RV^2_{span}} \sum\limits_{i=1}^N (RV_{i+1}-RV_i)^2\Bigg)$$
With N equal to the number of visits, and $RV_{span}=RV_{max}-RV_{min}$.}

\end{itemize}

    \subsection{\textit{Fitting procedure}}
    Radial velocities expected for each component were calculated from a model consisting of 6 parameters: velocity semi-amplitude of the primary ($K$), eccentricity ($e$), longitude of periastron ($\omega$), time of periastron ($T$), orbital period ($P$), and barycenter velocity ($\gamma$). A model radial velocity curve was computed, starting with the mean anomaly, $M$:
    $$M = \frac{2\pi}{P}(t-T)$$
    Using $M$, the eccentric anomaly, $E$ is computed:
    $$E = M + e \sin{(M)} + e^2 \frac{\sin{(2M)}}{2}$$
    Using $E$, the true anomaly $\nu$ is computed:
    $$\nu = 2 \arctan{\Bigg(\sqrt{\frac{1+e}{1-e}} \cdot \tan{\bigg(\frac{E}{2}\bigg)}\Bigg)}$$
    and finally the primary and secondary radial velocities were calculated:
    $$vel_{prim}=K[\cos{(\nu+\omega)}+e \cos{(\omega)}]+\gamma$$
    $$vel_{sec}=-\frac{K}{q}[\cos{(\nu+\omega)}+e \cos{(\omega)}]+\gamma$$
    $q$ was treated as a constant for each system, its value inherited from the method presented in \S 3.2

\par The set of orbital parameters $\bm{\theta} \equiv (K,e,\omega,T,P,\gamma)$ which accurately predicts the observed radial velocities of each component represents a possible orbital solution for the system.
    To find the best orbital fit we explored this parameter space using Bayesian techniques. We sampled the parameter space using \textit{emcee} \citep{Foreman-Mackey2013}, a Python implementation of an affine invariant ensemble sampler \citep{Goodman2010}. We used an ensemble of 4000 walkers, distributed evenly throughout the space, for 2000 steps. We kept the final 1000 steps of each run, discarding the initial 1000 as a burn-in phase.

\begin{figure*}[t!]
\centerline{
\includegraphics[scale=0.7]{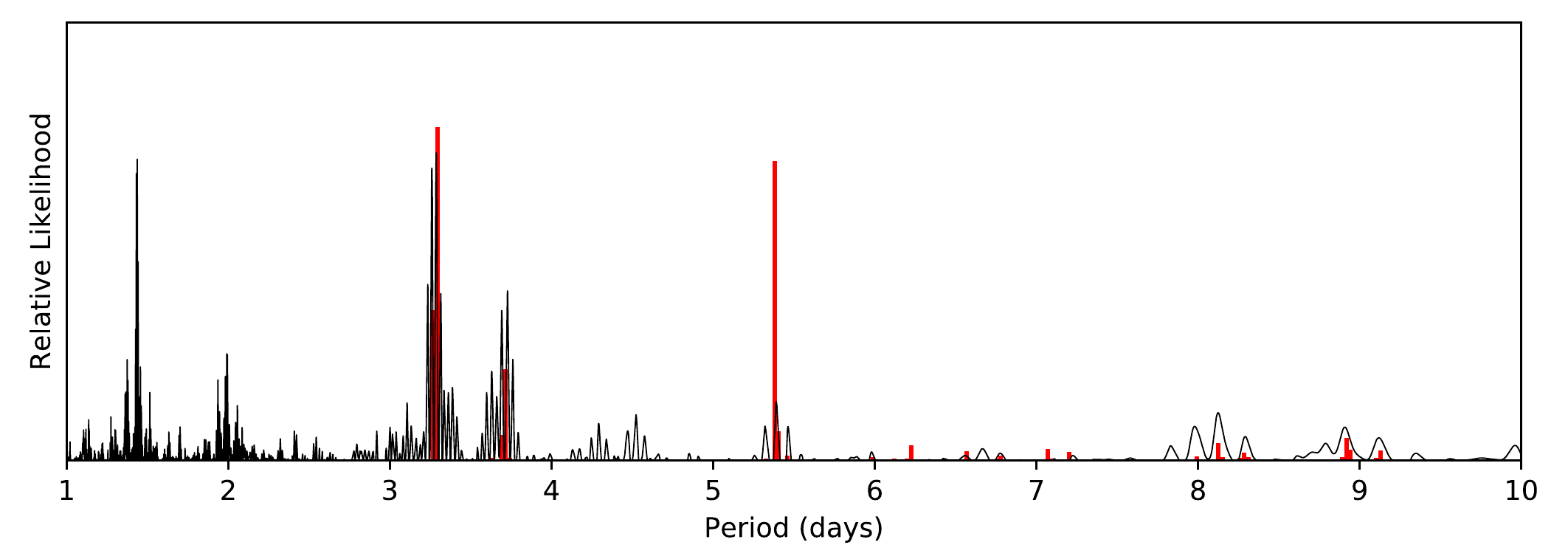}}
\caption{Periodogram power as a function of period (black), and a histogram of period samples drawn from the MCMC process (red) for 2M17204248+4205070.}
\label{fig:PeriodComparison}
\end{figure*}

\par For the number of visits typical of our orbital solutions (on the order of 10), the posterior probability distributions of $P$ were multimodal and highly degenerate. This made a period determination difficult. To perform the period search, we probed the parameter space using a modified likelihood function. The likelihood $p$ of observing the dataset $\bm{y}$ given $\bm{\theta}$ was defined:
$$p(\bm{y}|\bm{\theta})=\exp\Bigg[-\sqrt{\frac{1}{N}\sum\limits_{i=1}^N \frac{(o_i-c_i)^2}{\sigma_i}prim+\frac{(o_i-c_i)^2}{\sigma_i}sec}\Bigg]$$
where $o_i$ is the $i^{th}$ observation in $\bm{y}$ and $c_i$ is the computed radial velocity based on $\bm{\theta}$ at the $i^{th}$ epoch.
This definition prevents the ensemble from converging tightly on any single local maximum, allowing for multiple modes to be explored in a single walk. Figure \ref{fig:PeriodComparison} shows an example of the results of the MCMC period search using this likelihood definition, overlaid with a Lomb-Scargle periodogram. In Figure \ref{fig:PeriodComparison}, the samples are densest in period space at 3.29 days, corresponding to a peak in periodogram power. We define a period confidence, $\mathcal{L}$, as the fraction of MCMC samples contained within the primary peak identified by the period search: in the case shown in Figure \ref{fig:PeriodComparison}, $\mathcal{L}=30\%$. The highest period confidence value we measure is 79\%, for 2M21442066+4211363; for the other three systems, we measure period confidence values ranging from 16--56\%, suggesting that the maximum likelihood period is probable, but not yet definitively measured. The MCMC analysis also appears to favor shorter periods for these systems, producing a potential bias for other values inferred from the period measurements.
\par After constraining the period, we adopt the following priors for the other 5 parameters:

\begin{itemize}

\item{$0<K<100\frac{km}{s}$}
\item{$0<e<0.8$}
\item{$0<\omega<2\pi$}
\item{(median\,JD $-\frac{P}{2})<$T$<$ (median\,JD$+\frac{P}{2})$}
\item{min.\,observed\,RV$ <\gamma<$ max.\,observed\,RV}

\end{itemize}

In the random walk for the full orbital solution, we use the likelihood function:
    $$p(\bm{y}|\bm{\theta})=\exp\Bigg[-\sum\limits_{i=1}^N \frac{(o_i-c_i)^2}{2\sigma_i}prim+\frac{(o_i-c_i)^2}{2\sigma_i}sec\Bigg]$$
    This likelihood function reflects our assumption of independent, Gaussian probabilities.  For cases when the ensemble converged toward $e=0$, we performed a second run with a slightly modified circular orbit model, in which both $e$ and $\omega$ were constrained to 0. For each orbit fit, our choice of model was consistent with the Bayesian Information Criterion.
\par We report the $50^{th}$ quantile of the post burn-in distribution of the converged walkers as the value of each parameter. As uncertainties, we report the $16^{th}$ and $84^{th}$ quantiles as quasi $1\sigma$ values.
\par As a check against our orbital solutions, we estimate a lower bound on primary mass in Table \ref{table:orbit_fits} (See Appendix A for the derivation). All dynamical lower mass limits are significantly lower (by a factor of 5-10) than the photometric mass estimate of the primary listed in Table \ref{table:SB2s}. This indicates that either the sample includes an anomalously large number of high inclination systems, such that their dynamical mass is a significant underestimate of their true mass, or that the orbital period we have inferred is underestimated, as the dynamical mass estimate scales linearly with the system's assumed orbital period.  We suspect the latter explanation is more likely, and suggest that additional monitoring of these systems to remove the uncertainty that remains in the systems' periods is necessary.
    

\begin{deluxetable*}{cllll}
\tabletypesize{\scriptsize}
\tablecaption{Orbital fits}
\tablewidth{0pt}
\tablehead{\colhead{} & \colhead{2M06115599+3325505} & \colhead{2M17204248+4205070} & \colhead{2M21234344+4419277} & \colhead{2M21442066+4211363}}
\startdata
$K$ & $32.29\pm0.14$ & $43.87\pm0.08$ & $58.51\pm0.59$ & $61.16\pm0.46$\\
$e$ & $0.01\pm0.003$ & $0.002\pm0.002$ & $0.062\pm0.0.012$ & 0\\
$\omega$ & $128.94^{+30.75}_{-39.84}$ & $54.47^{+39.25}_{-25.77}$ & $127.78^{+8.69}_{-9.00}$ & 0\\
$T$ & $261.7467\pm0.006$ & $49.3840^{+0.3583}_{-0.2350}$ & $488.0751^{+0.1982}_{-0.2036}$ & $205.3381\pm0.0057$\\
$P(\mathcal{L})$ & $2.63(56\%)$ & $3.29(30\%)$ & $8.17(16\%)$ & $3.30(79\%)$\\
$\gamma$ & $76.98\pm0.06$ & $-6.77\pm0.05$ & $-123.59\pm0.45$ & $-17.22\pm0.37$\\
$\frac{M_{prim}}{M_{\odot}}$\tablenotemark{a}  & 0.005 & 0.017 & 0.089 & 0.04 \\
\enddata
\tablecomments{Units are $K(\frac{km}{s})$, $\omega$(degrees), $T$(JD$-$2456000), $P$(days), $\gamma(\frac{km}{s}$).}
\tablenotetext{a}{dynamical minimum mass estimates for the system's primary component, derived as explained in Appendix A. The minimum mass of an M dwarf primary is $M \geq 0.075 M_{\odot}$; individual photometric mass estimates for each primary are listed in Table \ref{table:SB2s}, with a range of 0.15-0.49 M$_{\odot}$.}
\label{table:orbit_fits}
\end{deluxetable*}


\section{RESULTS}

\textbf{Frequency: } As noted earlier, the systems which we detect as close multiples are a biased and incomplete subset of the larger, true population of close multiples within the parent sample of the SDSS-III/APOGEE M dwarf ancillary sample. Nonetheless,  it is instructive to compare the raw multiplicity fraction that we infer from this sample to prior measurements of the frequency of close companions to M-type primaries.  \citet{FischerMarcy1992} (FM1992) and \citet{Clark2012} (CBK2012) analyzed RV variability in multi-epoch optical spectra to infer a close ($a<$0.4~AU) binary fraction of 1.8$\pm$1.8\% and 3-4\%, respectively. More recently, \citet{Shan2015} analyzed the population of M+M eclipsing binaries in the Kepler field to infer a frequency of 11$\pm$2\% for close (P$<$90 days, or a$\leq$0.25 AU for a fiducial 0.5+0.5 M$_{\odot}$ system) companions to M dwarf primaries.

The raw (i.e., without corrections for incompleteness, inclination bias, etc.) binary fraction that we measure in our APOGEE sample ($\sim$3\%; 37/1350 = 2.74\%) seems to match the frequencies inferred by FM1992 and CBK2012. Those prior measurements have been corrected for incompleteness and selection effects, however, while our raw multiplicity fraction has not. The only selection effect that would drive our raw multiplicity fraction to overestimate the true value is the magnitude limited nature of the APOGEE M dwarf ancillary targets; biases due to inclination, flux ratios, and temporal sampling will all conspire to make our empirically measured rate underestimate the true multiplicity fraction. Thus, our measurement should be an underestimate of the true intrinsic multiplicity fraction. If completeness correction shows that our current binary detection rate is above 50\%, then our result would support that of FM1992 and CBK2012. If our current binary detection rate is below 50\%, then our result would more strongly favor \citet{Shan2015}. Detailed modeling will be necessary to draw strong conclusions, and should be the subject of future work.

\textbf{Mass Ratio Distribution: } Measuring the mass ratios of 29 M Dwarf SB2s and 5 SB3s, we find a mass ratio distribution that reaches as low as 0.217, but with most systems having mass ratios between 0.8 and 1.  As noted in section \ref{sec:detection_limits}, requiring the detection of spectral counterparts for both components of the system will bias our sample towards equal mass ratios. Nonetheless, it is again instructive to compare our mass ratio distribution to those measured in existing samples of SB2s, particularly as those catalogs will suffer from similar selection biases. Our cumulative mass ratio distribution shows fair agreement to those found by \citet{Pourbaix2004} and \citet{Fernandez2017}: while our mass ratio distribution appears somewhat more strongly skewed towards equal mass systems, a Kolmogorov-Smirnov test finds a $\sim$18\% chance that the mass ratio distribution that we measure for the APOGEE M dwarf SB2s is consistent with those measured \citet{Pourbaix2004} and \citet{Fernandez2017} for similarly biased catalogs of (higher mass) SB2s. 
 
\begin{figure*}
\centering
\includegraphics[scale=.5]{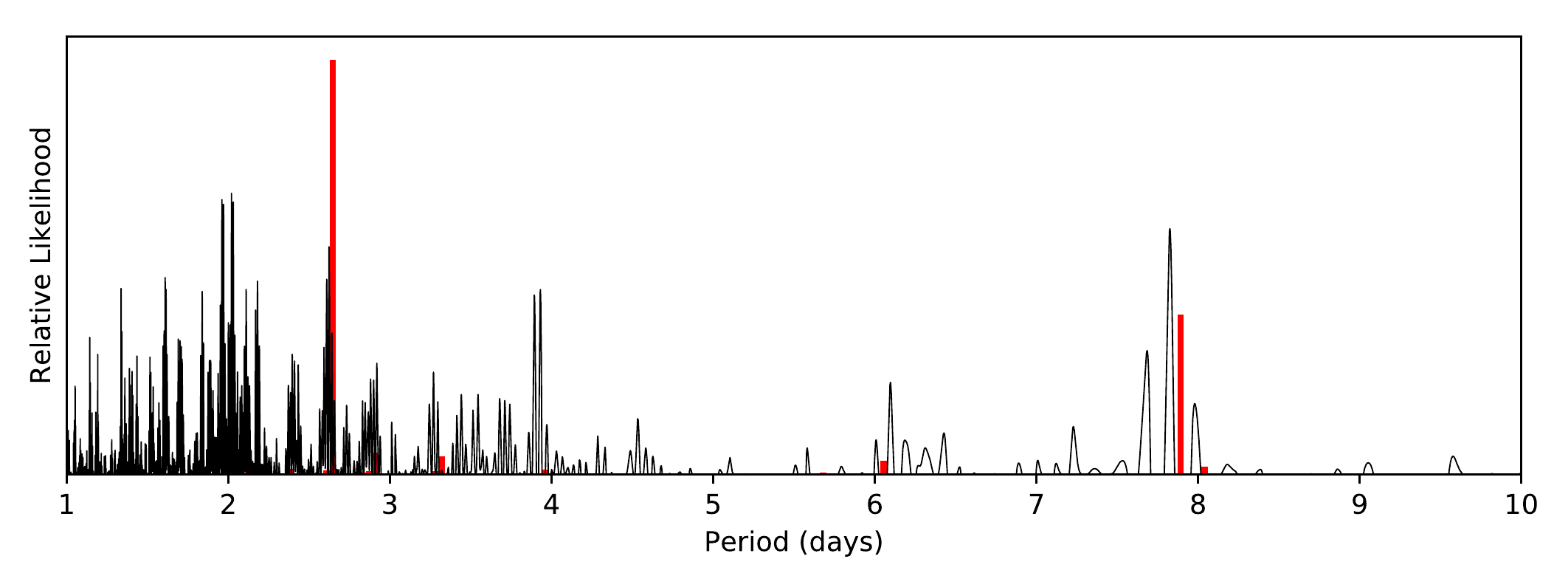}
\centerline{
\includegraphics[scale=.25]{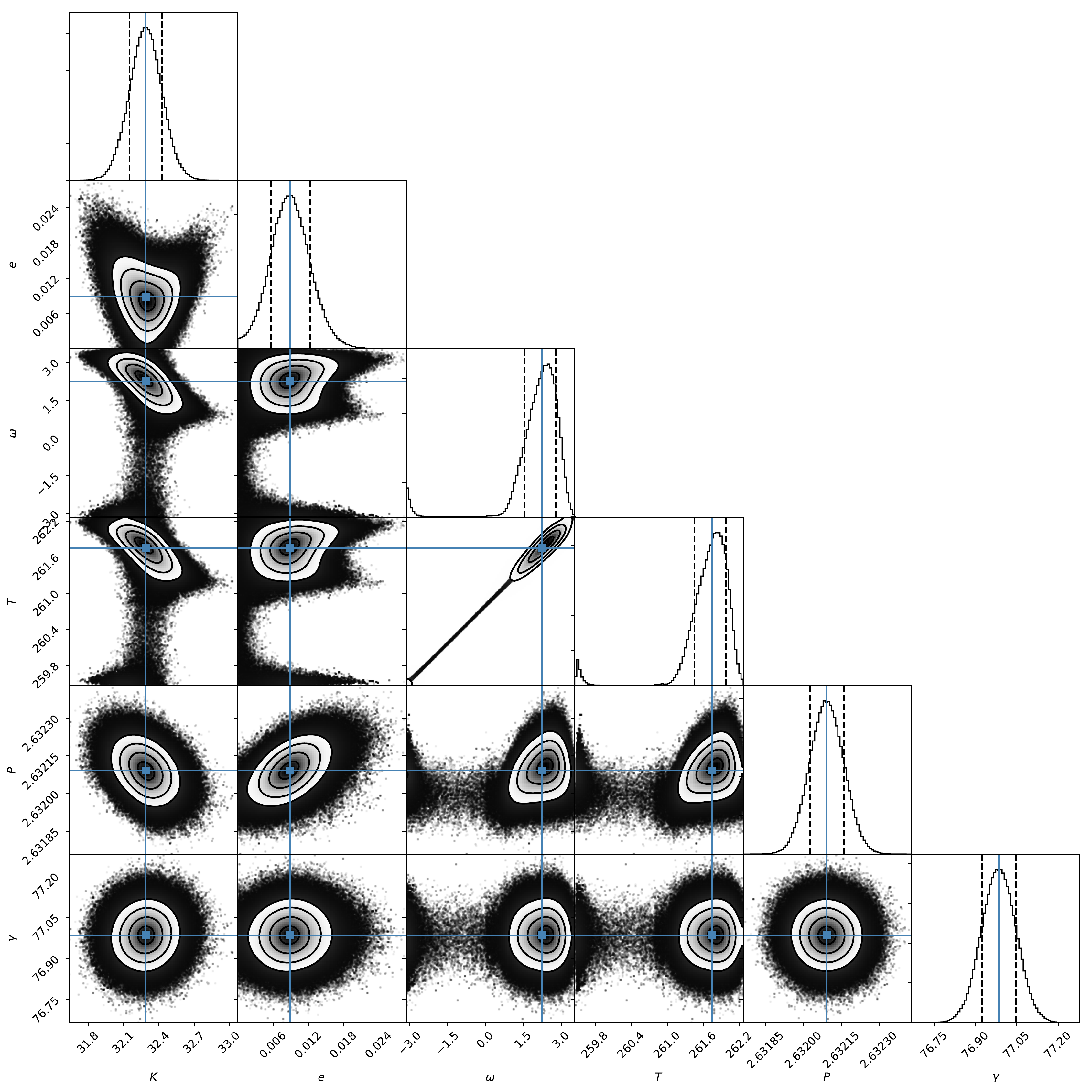}
\includegraphics[scale=.35]{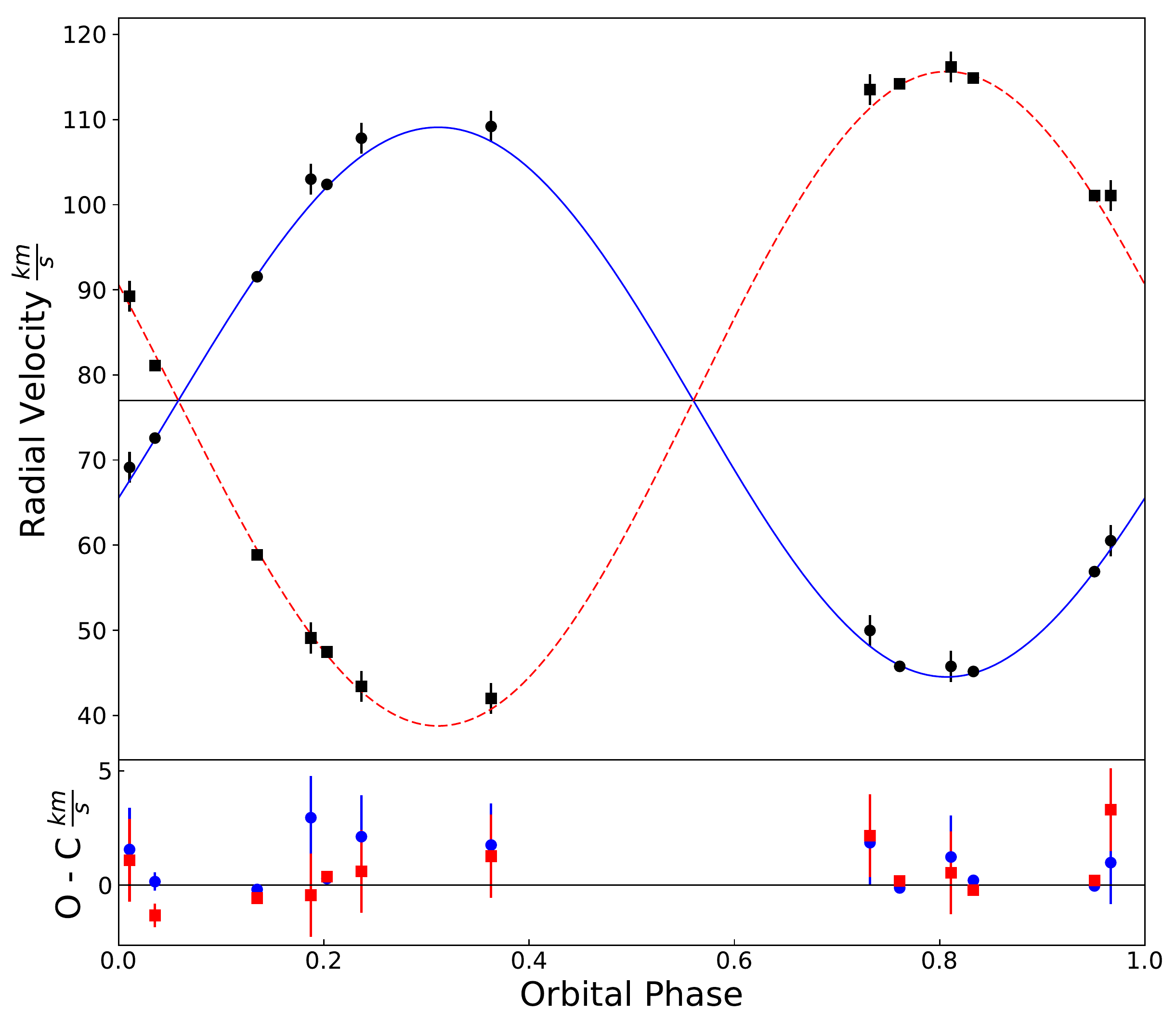}}
\caption{Figures for 2M06115599+3325505. $\mathcal{L}=56\%$ at $P=2.63$ days.
\newline [Top] Lomb-Scargle Periodogram power as a function of period (black), and the histogram of MCMC samples obtained during the period search (red).
\newline [Left] Corner plot of the posterior probability distribution given by the random walk.
\newline [Right] Radial velocity plot of this system. The solid blue curve is the primary component velocity, and the red dashed curve is the secondary component velocity. Primary component velocities are marked as squares, secondary velocities are marked as circles.}
\label{fig:0611+3325}
\end{figure*}

\begin{figure*}
\centering
\includegraphics[scale=.5]{4205_period_plot.pdf}
\centerline{
\includegraphics[scale=.25]{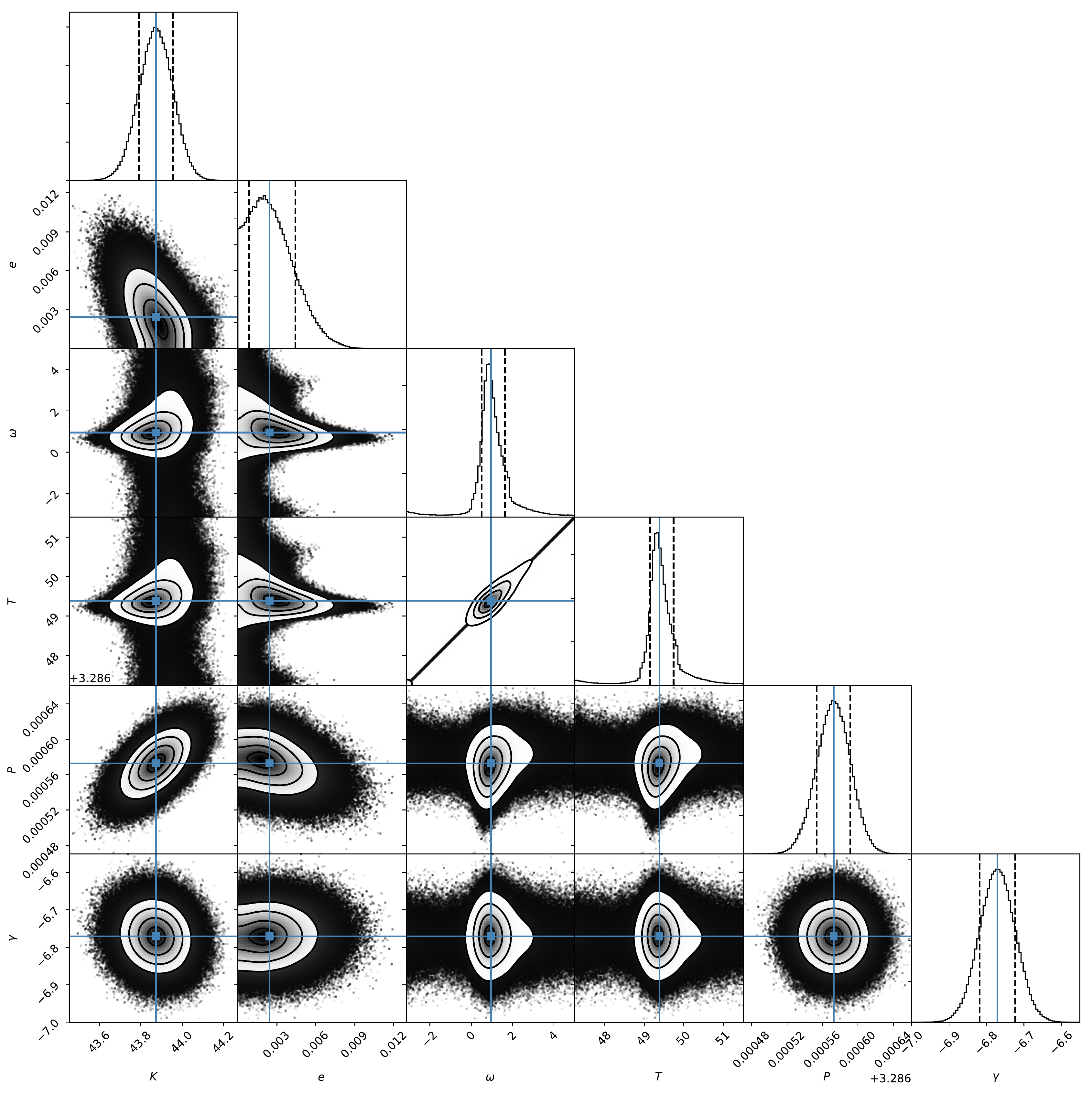}
\includegraphics[scale=.35]{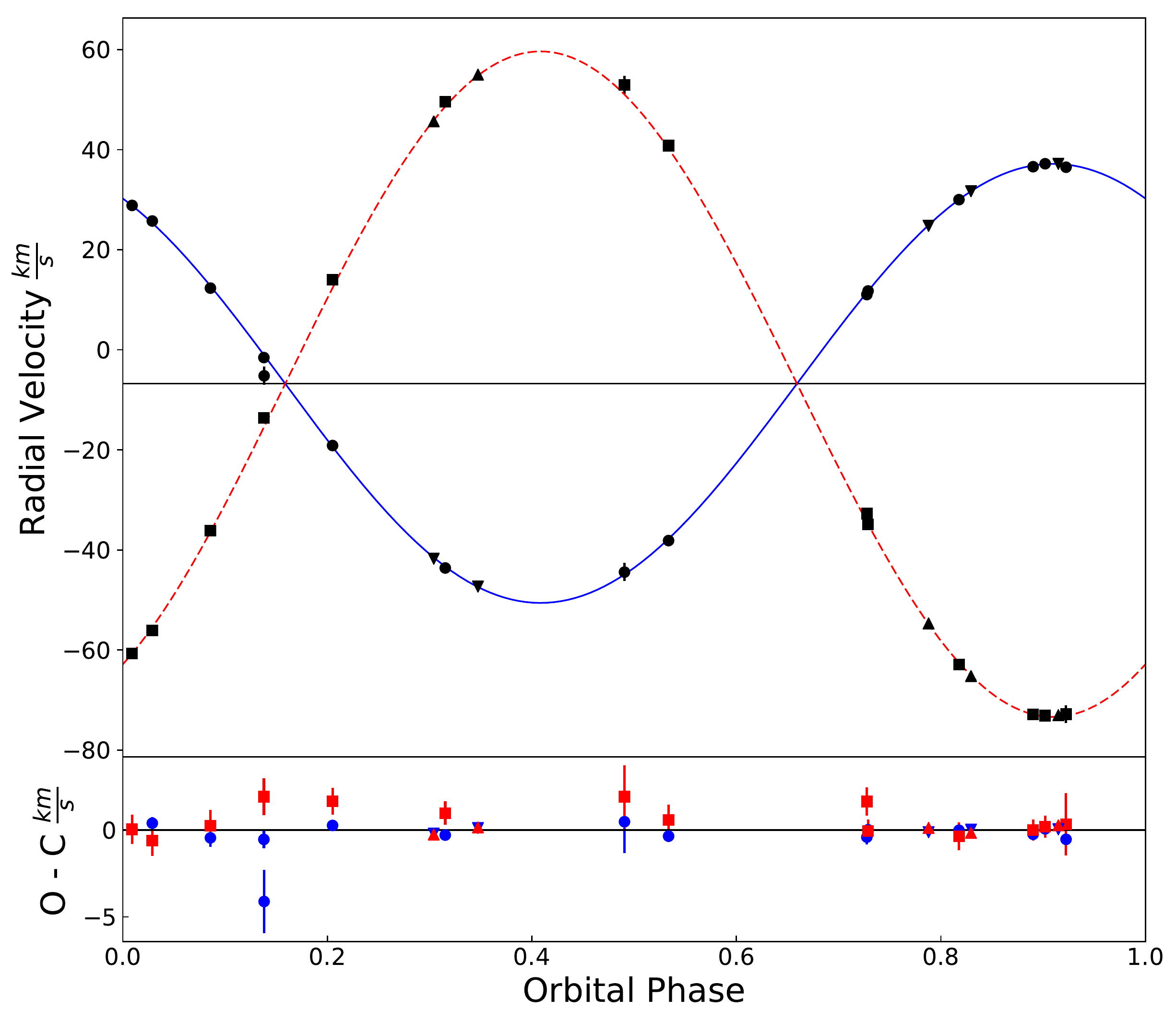}}
\caption{Figures for 2M17204248+4205070. $\mathcal{L}=30\%$ at $P=3.29$ days.
\newline [Top] Lomb-Scargle Periodogram power as a function of period (black), and the histogram of MCMC samples obtained during the period search (red).
\newline [Left] Corner plot of the posterior probability distribution given by the random walk.
\newline [Right] Radial velocity plot of this system. The solid blue curve is the primary component velocity, and the red dashed curve is the secondary component velocity. Primary component velocities are marked as squares, secondary velocities are marked as circles.}
\label{fig:1720+4205}
\end{figure*}

\begin{figure*}
\centering
\includegraphics[scale=.5]{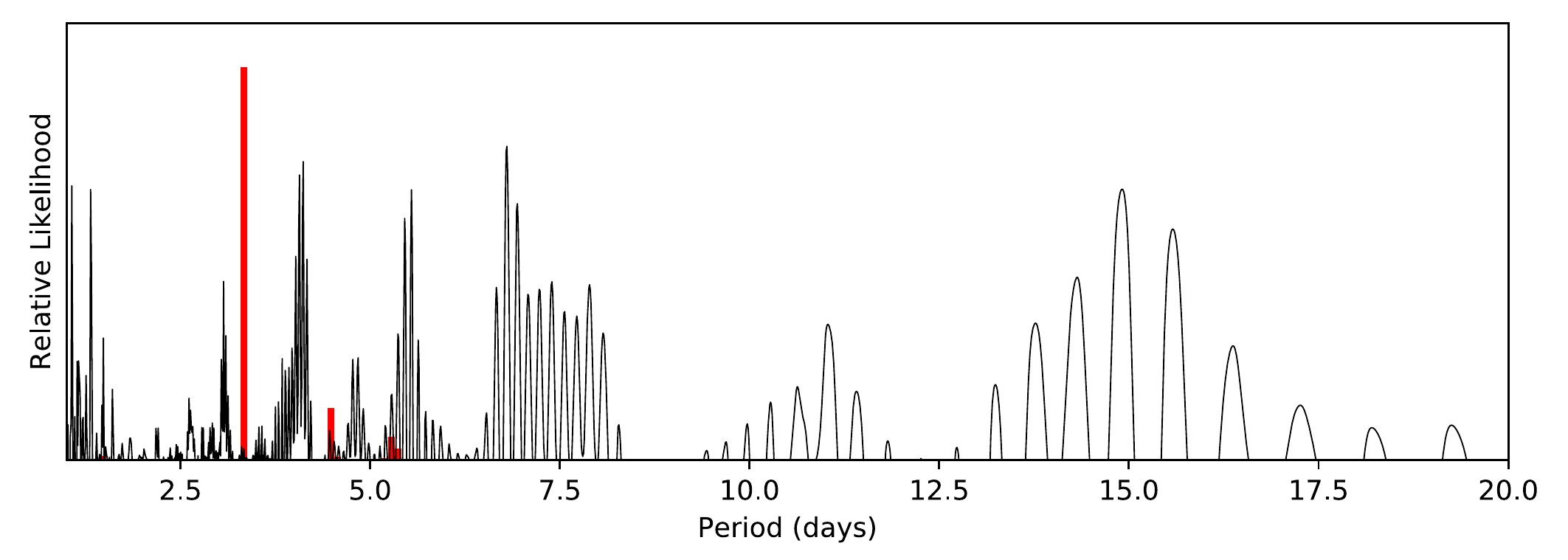}
\centerline{
\includegraphics[scale=.35]{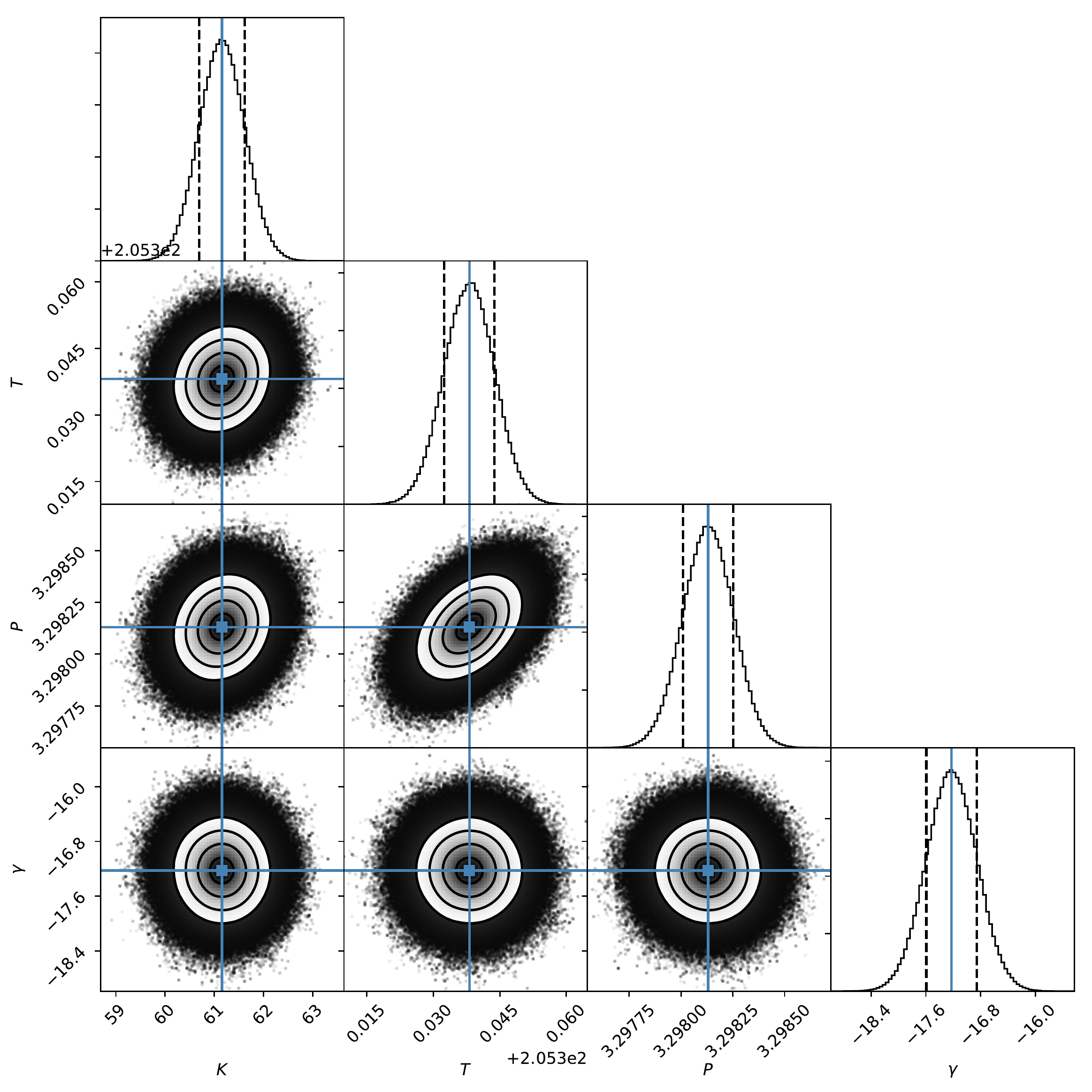}
\includegraphics[scale=.35]{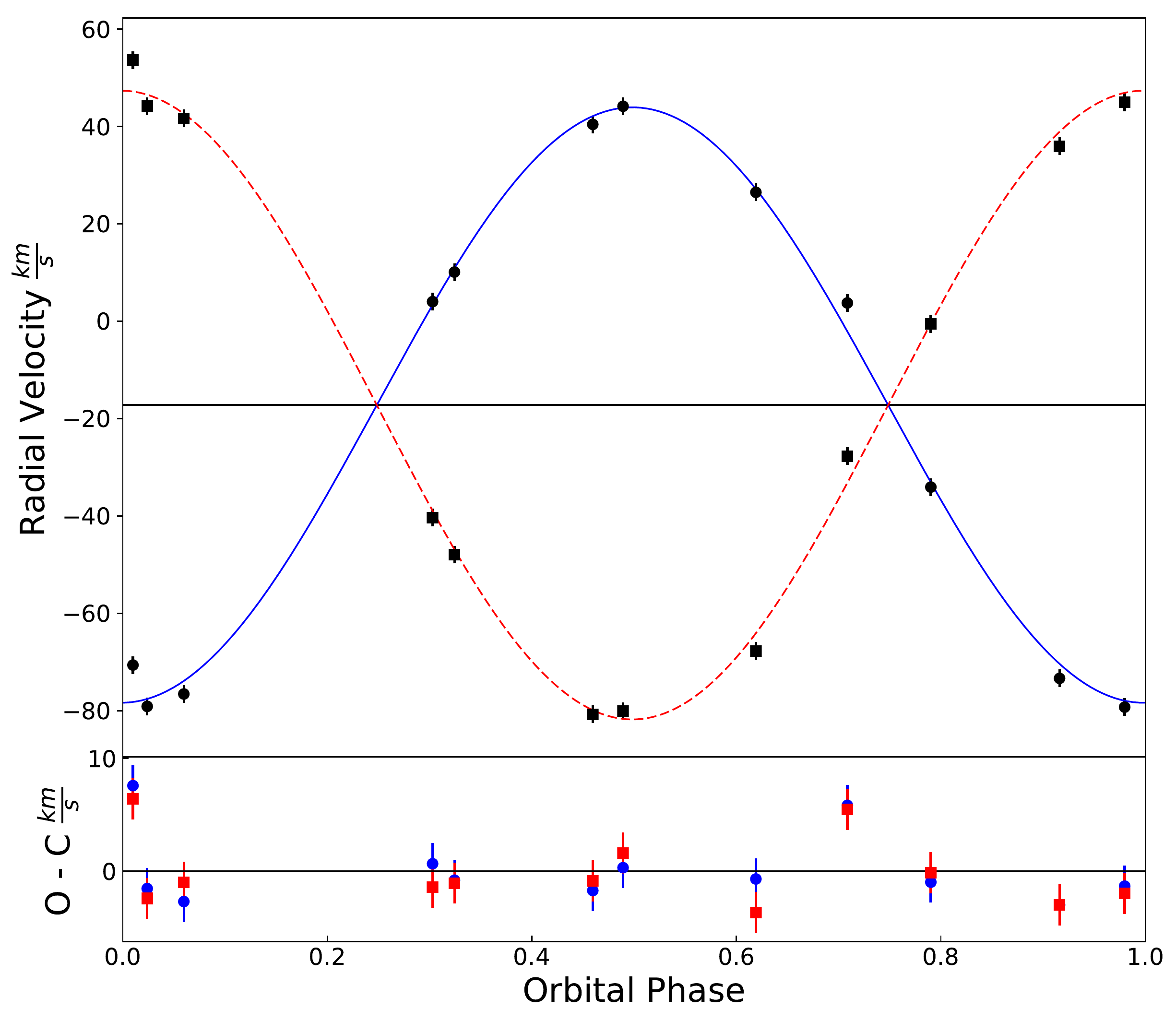}}
\caption{Figures for 2M21442066+4211363. $\mathcal{L}=79\%$ at $P=3.30$ days.
\newline [Top] Lomb-Scargle Periodogram power as a function of period (black), and the histogram of MCMC samples obtained during the period search (red).
\newline [Left] Corner plot of the posterior probability distribution given by the random walk.
\newline [Right] Radial velocity plot of this system. The solid blue curve is the primary component velocity, and the red dashed curve is the secondary component velocity. Primary component velocities are marked as squares, secondary velocities are marked as circles.}
\label{fig:2144+4211}
\end{figure*}

\textbf{Orbits: } Orbital fit results are tabulated in Table \ref{table:orbit_fits}. All 4 orbital solutions exhibit small eccentricities ($e<0.1$). We find periods of 2.6-8.2 days. In the context of a multimodal period distribution, defining uncertainty via peak width is problematic. As a result, we omit uncertainties on our measurement of orbital period in favor of the period confidence defined in \S4.2. Three key figures (see Figures \ref{fig:0611+3325}-\ref{fig:2144+4211}) are included for each orbital solution, following \S6.

\par

\section{CONCLUSIONS}

\begin{enumerate}
\item{We have identified 44 candidate close multiple systems among the 1350 targets in the SDSS-III/APOGEE M dwarf ancillary sample. These candidates include 8 of the 9 SB2s previously identified by \citet{Deshpande2013} in their analysis of a subset of the APOGEE M dwarf sample, as well as 3 SB2s and an SB3 identified by \citet{ElBadry2018} in their search for binaries within DR14, indicating that our algorithm successfully recovers close binaries whose APOGEE spectra capture an epoch where the system exhibits a large velocity separation.}
\item{We have extracted RVs for components in 34 of these systems, including 5 systems that appear to be higher order multiples.  In most cases, these RV measurements are obtained by fitting peaks in the CCFs produced by the APOGEE reduction and analysis pipeline; in systems with more than two components, or with velocity separations too small to resolve in the APOGEE CCF, we have extracted RVs using the TODCOR algorithm on the APOGEE spectra themselves. For two stars, we have also obtained follow-up spectroscopy with the High Resolution Spectrograph on the Hobby-Eberly Telescope; we analyze these optical/far-red data using the TODCOR routine as well.}
\item{We fit primary and secondary RVs to measure mass ratios for the closest pair of each of the 34 systems for which we extract RVs. The mass ratio distribution of close pairs in our sample skews towards equal mass systems, and includes only one system with a mass ratio $<$ 0.45; this is consistent with first order estimates of the bias towards higher mass ratios that should result from requiring a positive spectroscopic detection of both primary and secondary components.  Nonetheless, the (biased and incomplete) mass ratio distribution that we measure from the M dwarf sample is consistent at the 1$\sigma$ level with the mass ratio distributions reported in the literature for similarly biased samples of younger and more massive stars, suggesting that the mass ratios of close multiples are not a strong function of primary mass.}
\item{The low periods we measure for our targets ($P<10$ days) are consistent with the small separations we expect for M Dwarf SB2s. The low eccentricities we measure ($e<0.1$) reflect the tidal interactions to which close binaries are subject. Our orbit fits exhibit small residuals, excluding third bodies down to very low masses.}
\end{enumerate}

\section*{ACKNOWLEDGEMENTS}

J.S., K.R.C., and M.K. acknowledge support provided by the NSF through grant AST-1449476, and from the Research Corporation via a Time Domain Astrophysics Scialog award. C.F.B.\ and N.R.\ acknowledge support provided by the NSF through grant AST-1517592. N.D. acknowledges support for this work from the NSF through grant AST-1616684.

Funding for SDSS-III has been provided by the Alfred P. Sloan Foundation, the Participating Institutions, the National Science Foundation, and the U.S. Department of Energy Office of Science. The SDSS-III web site is http://www.sdss3.org/.

SDSS-III is managed by the Astrophysical Research Consortium for the Participating Institutions of the SDSS-III Collaboration including the University of Arizona, the Brazilian Participation Group, Brookhaven National Laboratory, Carnegie Mellon University, University of Florida, the French Participation Group, the German Participation Group, Harvard University, the Instituto de Astrofisica de Canarias, the Michigan State/Notre Dame/JINA Participation Group, Johns Hopkins University, Lawrence Berkeley National Laboratory, Max Planck Institute for Astrophysics, Max Planck Institute for Extraterrestrial Physics, New Mexico State University, New York University, Ohio State University, Pennsylvania State University, University of Portsmouth, Princeton University, the Spanish Participation Group, University of Tokyo, University of Utah, Vanderbilt University, University of Virginia, University of Washington, and Yale University.

D.A.G.H. was funded by the Ram\'on y Cajal fellowship number RYC-2013-14182.
D.A.G.H. and O.Z. acknowledge support provided by the Spanish Ministry of
Economy and Competitiveness (MINECO) under grant AYA-2014-58082-P.

\newpage

\newpage

\begin{figure*}
\centering
\includegraphics[scale=.5]{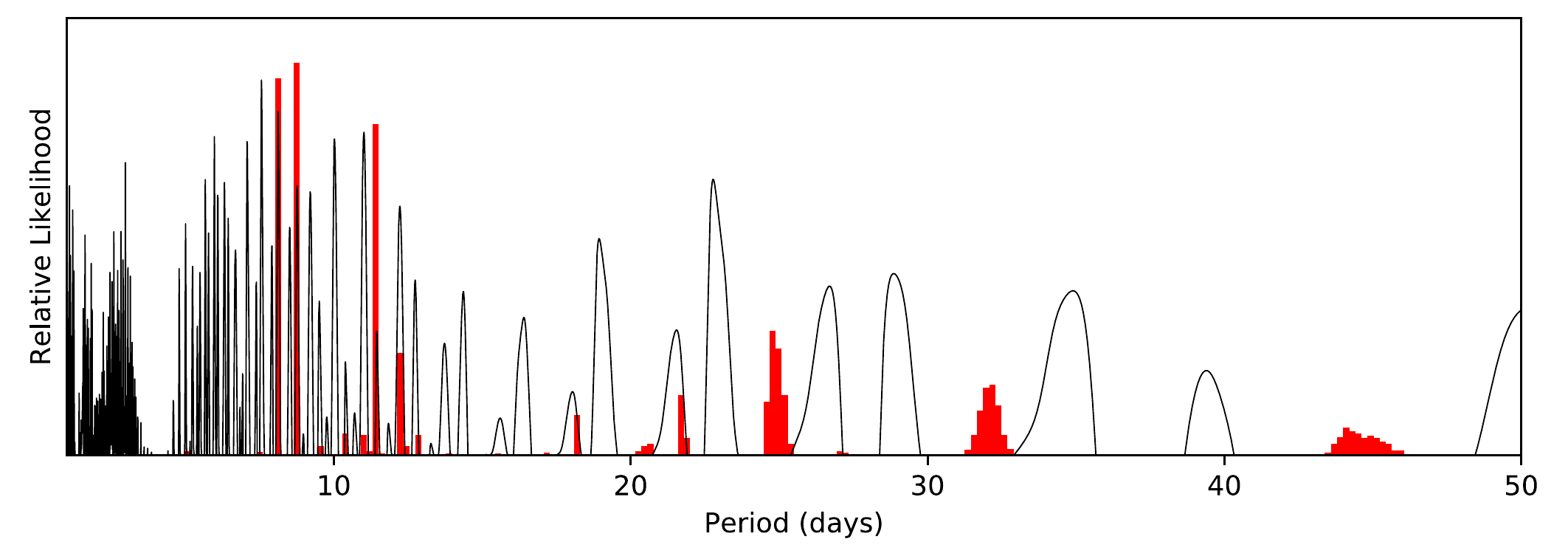}
\centerline{
\includegraphics[scale=.25]{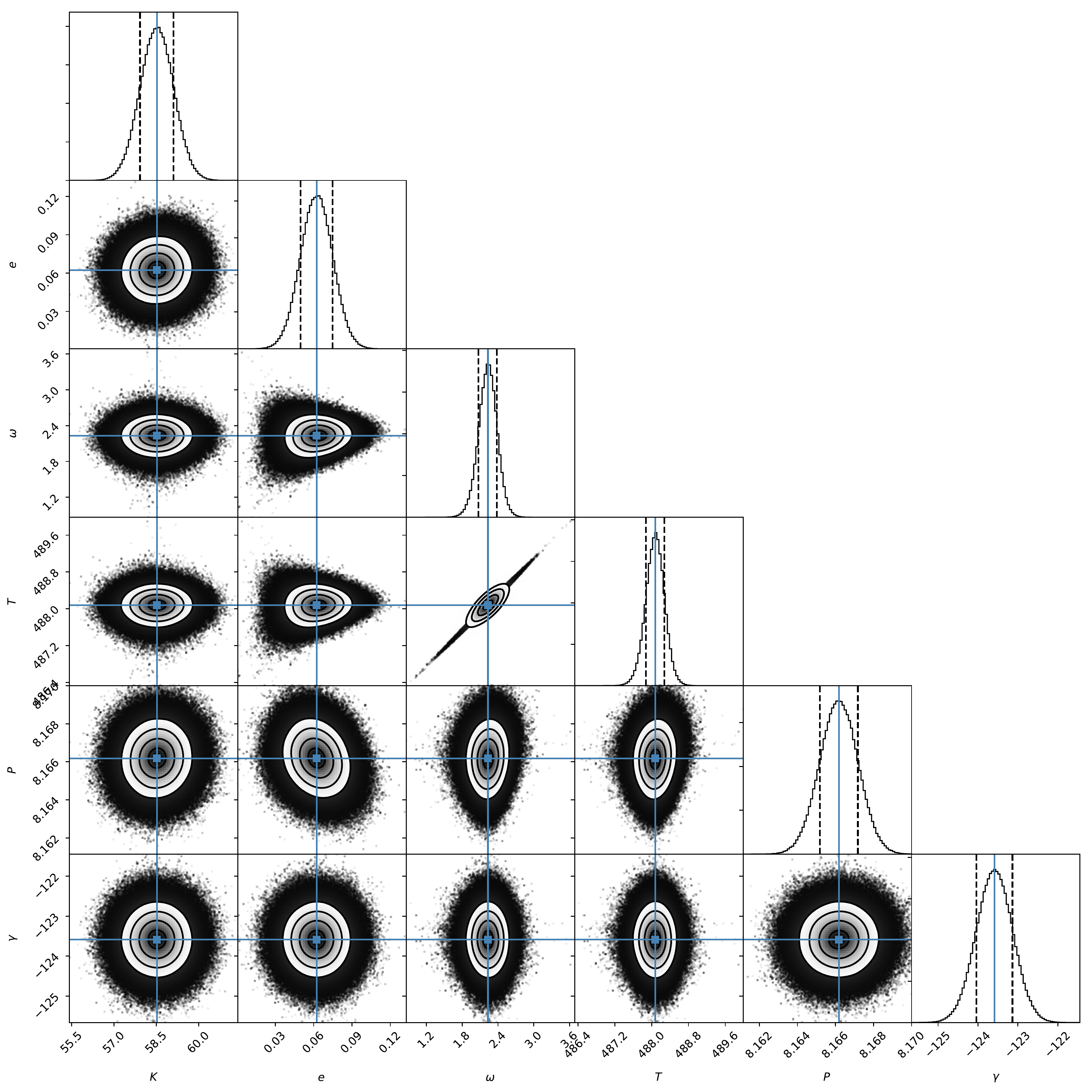}
\includegraphics[scale=.35]{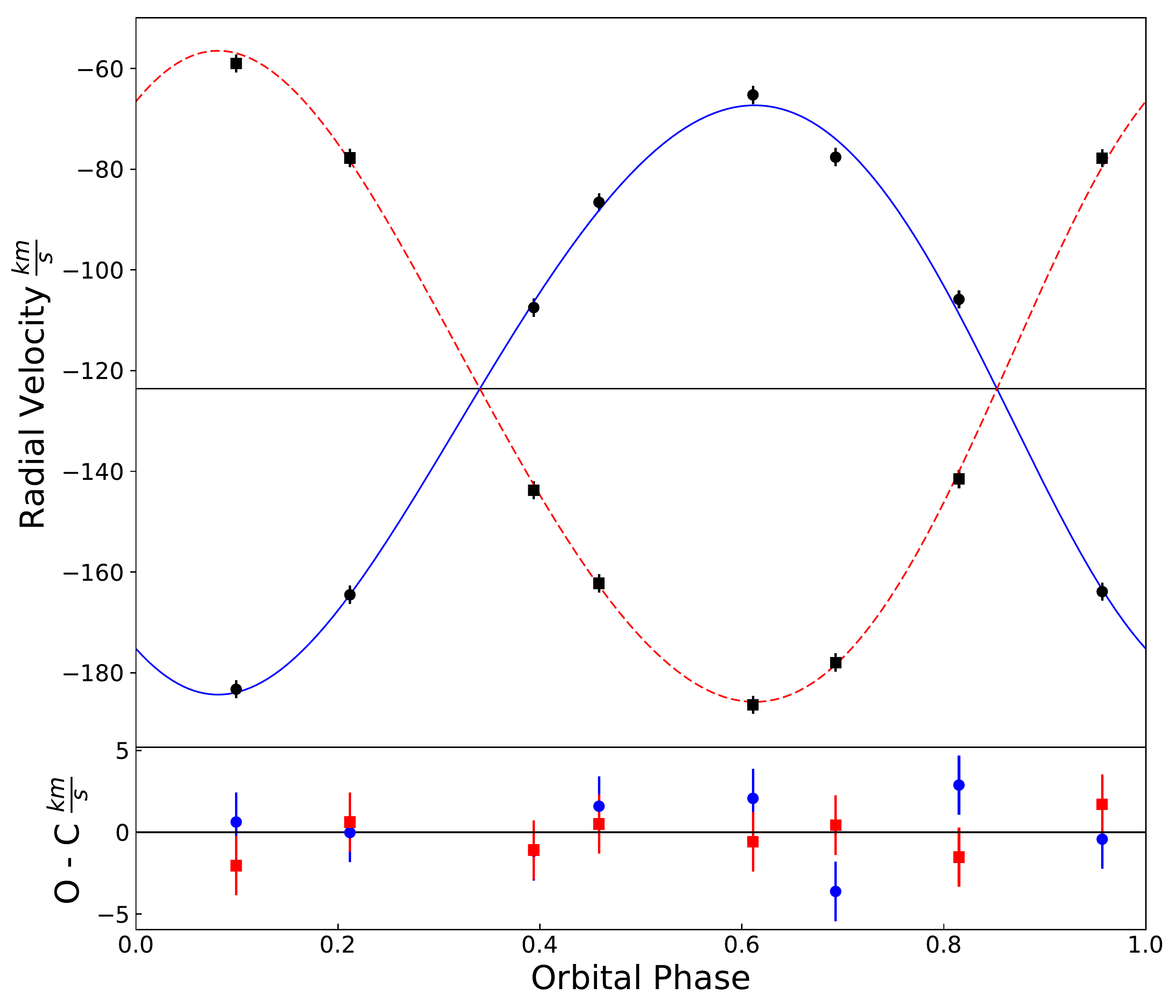}}
\caption{Figures for 2M21234344+4419277. $\mathcal{L}=16\%$ at $P=8.17$ days.
\newline [Top] Lomb-Scargle Periodogram power as a function of period (black), and the histogram of MCMC samples obtained during the period search (red).
\newline [Left] Corner plot of the posterior probability distribution given by the random walk.
\newline [Right] Radial velocity plot of this system. The solid blue curve is the primary component velocity, and the red dashed curve is the secondary component velocity. Primary component velocities are marked as squares, secondary velocities are marked as circles.}
\end{figure*}

\newpage
        
\clearpage

\appendix{}
\section{Estimate of Primary Mass Lower Bound}
Given that all stars in our sample are known to be M Dwarfs from their spectral information, we know that any calculated lower limit on the mass of the primary component must not fall above about 0.5$M_{\odot}$. This calculation acts as an independent check on the orbital fit, which is useful given the solution degeneracy that is prevalent along the period axis.
\newline
\newline Isolate $P^2$ from Newton's formulation of Kepler's $3^{rd}$ law

$$ P^2 = (2\pi)^2  \frac{a^3}{GM_{prim}(1+q)}$$

and an approximate expression of mean orbital speed

$$v_o \approx \frac{2\pi a}{P}\left[1-\frac{1}{4}e^2-\frac{3}{64}e^4 \right]$$

where $v_o \approx K$. This approximation is sufficiently accurate for low $e$, and pushes the estimate down for non-zero $e$. Edge-on orientation is assumed, firmly establishing this calculation as a lower bound on primary mass.

$$K \leq \frac{2\pi a}{P}\left[1-\frac{1}{4}e^2-\frac{3}{64}e^4 \right]$$

substitution and simplification ultimately gives

$$  M_{prim} \geq \frac{K^3P}{2\pi G(1+q)\left[1-\frac{1}{4}e^2-\frac{3}{64}e^4 \right]^3}$$

If the orbital solution places this value significantly above 0.5$M_{\odot}$, then it cannot be a correct 
solution.

\pagebreak

\iftrue 
\section{Data Tables}

The list of SB2 Radial Velocity measurements is given here in its entirety.

\startlongtable
\begin{deluxetable*}{crccrrr}
\tabletypesize{\scriptsize}
\tablecaption{Radial Velocity Measurements of SB2s}
\tablewidth{0pt}
\tablehead{
\colhead{2MASS ID} & \colhead{Visit} & \colhead{Epoch (MJD)} & \colhead{SDSS plate \& Fiber} & \colhead{SNR} & \colhead{v$_{prim}(\frac{km}{s})$} & \colhead{v$_{sec}(\frac{km}{s})$}}
\startdata
2M03393700+4531160 & 1 & 56195.3409 & 6244-56195-086 & 117 & -16.2 & 33.8 \\
$\vert$ & 2 & 56200.2983 & 6244-56200-131 & 210 & -20.4 & 38.9 \\
$\vert$ & 3 & 56223.2868 & 6244-56223-131 & 215 & -37.3 & 53.8 \\
$\vert$ & 4 & 56196.3190 & 6245-56196-077 & 168 & 52.0 & -36.3 \\
$\vert$ & 5 & 56202.2755 & 6245-56202-074 & 137 & 7.5 & - \\
$\vert$ & 6 & 56224.3188 & 6245-56224-077 & 186 & 11.1 & 3.9 \\
2M04373881+4650216 & 1 & 56176.4835 & 6212-56176-050 & 49 & -40.8 & -44.4 \\
$\vert$ & 2 & 56234.3042 & 6212-56234-050 & 32 & -31.5 & -56.4 \\
$\vert$ & 3 & 56254.2442 & 6212-56254-050 & 62 & -26.1 & -63.8 \\
$\vert$ & 4 & 56260.2176 & 6212-56260-050 & 44 & -26.6 & -61.6 \\
2M05421216+2224407 & 1 & 56291.1826 & 6761-56291-131 & 25 & 10.0 & 49.2 \\
$\vert$ & 2 & 56559.4916 & 6761-56559-134 & 78 & 6.3 & 48.0 \\
$\vert$ & 3 & 56582.4342 & 6761-56582-128 & 88 & 48.6 & 7.1 \\
$\vert$ & 4 & 56586.4264 & 6761-56586-128 & 69 & 48.4 & 7.6 \\
2M05504191+3525569 & 1 & 56262.3018 & 6543-56262-287 & 95 & 83.0 & 69.1 \\
$\vert$ & 2 & 56285.1795 & 6543-56285-284 & 37 & 73.9 & 88.9 \\
$\vert$ & 3 & 56313.1350 & 6543-56313-296 & 80 & 95.1 & 57.8 \\
2M06115599+3325505 & 1\tablenotemark{1} & 55848.4441 & 5508-55848-274 & 157 & 45.16$\pm0.19$ & 114.89$\pm0.25$ \\
$\vert$ & 2\tablenotemark{1} & 55849.4183 & 5507-55849-220 & 131 & 102.38$\pm0.18$ & 47.46$\pm0.23$ \\
$\vert$ & 3\tablenotemark{1} & 55927.2179 & 5507-55927-226 & 110 & 45.76$\pm0.19$ & 114.19$\pm0.25$ \\
$\vert$ & 4\tablenotemark{1} & 55928.2021 & 5508-55928-219 & 181 & 91.53$\pm0.17$ & 58.87$\pm0.21$ \\
$\vert$ & 5\tablenotemark{1} & 55933.2042 & 5507-55933-219 & 161 & 72.58$\pm0.40$ & 81.08$\pm0.52$ \\
$\vert$ & 6\tablenotemark{1} & 55967.1987 & 5508-55967-220 & 150 & 56.89$\pm0.17$ & 101.07$\pm0.23$ \\
$\vert$ & 7 & 56260.4457 & 6344-56260-226 & 202 & 109.2 & 42.0 \\
$\vert$ & 8 & 56267.4147 & 6344-56267-225 & 157 & 69.1 & 89.3 \\
$\vert$ & 9 & 56288.3556 & 6344-56288-225 & 21 & 60.5 & 101.1 \\
$\vert$ & 10 & 56314.2669 & 6344-56314-225 & 200 & 45.8 & 116.2 \\
$\vert$ & 11 & 56290.3703 & 6345-56290-219 & 153 & 50.0 & 113.5 \\
$\vert$ & 12 & 56315.2574 & 6345-56315-220 & 161 & 103.0 & 49.1 \\
$\vert$ & 13 & 56323.2832 & 6345-56323-220 & 144 & 107.8 & 43.4 \\
2M06125378+2343533 & 1 & 56347.1522 & 6548-56347-027 & 94 & 1.0 & -41.5 \\
$\vert$ & 2 & 56352.1825 & 6548-56352-021 & 88 & 1.8 & -41.1 \\
$\vert$ & 3 & 56623.4989 & 6548-56623-015 & 95 & -35.7 & 5.3 \\
2M06213904+3231006 & 1 & 56260.4457 & 6344-56260-084 & 85 & 50.9 & -23.4 \\
$\vert$ & 2 & 56267.4147 & 6344-56267-084 & 65 & 49.3 & -21.6 \\
$\vert$ & 3 & 56314.2669 & 6344-56314-031 & 66 & 14.9 & - \\
$\vert$ & 4 & 56290.3703 & 6345-56290-080 & 56 & 50.5 & -23.3 \\
$\vert$ & 5 & 56315.2574 & 6345-56315-065 & 62 & 58.1 & -28.8 \\
$\vert$ & 6 & 56323.2832 & 6345-56323-062 & 49 & 37.2 & -10.5 \\
2M06561894-0835461 & 1 & 56587.4858 & 6535-56587-178 & 244 & 0.4 & - \\
$\vert$ & 2 & 56593.4881 & 6535-56593-171 & 199 & -36.9 & - \\
$\vert$ & 3 & 56616.3926 & 6535-56616-159 & 68 & -29.9 & 53.0 \\
$\vert$ & 4 & 56637.3386 & 6535-56637-154 & 217 & 26.7 & -38.6 \\
2M07063543+0219287 & 1\tablenotemark{2} & 56673.2300 & 6552-56673-250 & 229 & 85.72 & 66.91 \\
$\vert$ & 2 & 56677.2508 & 6552-56677-250 & 232 & 102.5 & 44.5 \\
$\vert$ & 3\tablenotemark{2} & 56700.2056 & 6552-56700-250 & 245 & 63.09 & 79.31 \\
2M07444028+7946423 & 1 & 56349.2512 & 6566-56349-237 & 213 & 8.6 & -43.7 \\
$\vert$ & 2 & 56640.4901 & 6566-56640-279 & 232 & -26.6 & -14.4 \\
$\vert$ & 3 & 56650.4149 & 6566-56650-226 & 176 & -66.7 & 24.4 \\
2M08100405+3220142 & 1\tablenotemark{1} & 56323.3399 & 6778-56323-213 & 143 & 7.24$\pm0.15$ & 30.69$\pm0.15$ \\
$\vert$ & 2\tablenotemark{1} & 56349.3061 & 6778-56349-225 & 163 & 21.84$\pm0.16$ & 14.99$\pm0.18$ \\
$\vert$ & 3\tablenotemark{1} & 56353.2618 & 6778-56353-213 & 167 & 18.79$\pm0.22$ & 18.56$\pm0.23$ \\
$\vert$ & 4\tablenotemark{1} & 56324.3037 & 6779-56324-220 & 145 & 6.50$\pm0.19$ & 31.19$\pm0.20$ \\
$\vert$ & 5\tablenotemark{1} & 56352.2445 & 6779-56352-220 & 144 & 18.73$\pm0.25$ & 18.62$\pm0.27$ \\
$\vert$ & 6\tablenotemark{1} & 56372.2128 & 6779-56372-220 & 137 & 35.69$\pm0.13$ & 1.68$\pm0.14$ \\
2M10423925+1944404 & 1\tablenotemark{1} & 56264.5342 & 6576-56264-045 & 57 & -4.42$\pm0.40$ & -25.49$\pm0.40$ \\
$\vert$ & 2\tablenotemark{1} & 56288.5058 & 6613-56288-052 & 87 & -1.96$\pm0.32$ & -28.09$\pm0.32$ \\
$\vert$ & 3\tablenotemark{1} & 56313.3796 & 6613-56313-051 & 67 & -0.56$\pm0.47$ & -30.12$\pm0.44$ \\
$\vert$ & 4\tablenotemark{1} & 56314.3304 & 6613-56314-051 & 72 & -1.59$\pm0.39$ & -29.68$\pm0.43$ \\
2M10464238+1626144 & 1\tablenotemark{1} & 55967.3129 & 5680-55967-015 & 56 & -15.51$\pm0.32$ & -58.44$\pm0.47$ \\
$\vert$ & 2\tablenotemark{1} & 55989.2456 & 5680-55989-022 & 43 & -66.65$\pm0.26$ & 1.45$\pm0.43$ \\
$\vert$ & 3\tablenotemark{1} & 55990.2372 & 5680-55990-022 & 72 & -22.06$\pm0.24$ & -52.50$\pm0.35$ \\
2M11081979+4751217 & 1 & 56675.4195 & 7348-56675-089 & 83 & -30.4 & 31.1 \\
$\vert$ & 2 & 56698.3212 & 7348-56698-089 & 56 & - & - \\
$\vert$ & 3 & 56700.3171 & 7348-56700-089 & 66 & -30.8 & 33.5 \\
$\vert$ & 4 & 56704.3391 & 7348-56704-071 & 8 & -29.9 & 28.1 \\
$\vert$ & 5 & 56729.2837 & 7348-56729-065 & 69 & -28.3 & 21.8 \\
2M12045611+1728119 & 1\tablenotemark{1} & 55968.3811 & 5683-55968-260 & 205 & -46.36$\pm0.16$ & 95.09$\pm0.48$ \\
$\vert$ & 2\tablenotemark{1} & 56022.2465 & 5683-56022-260 & 173 & -19.67$\pm0.22$ & 53.37$\pm0.59$ \\
$\vert$ & 3 & 56379.2677 & 5683-56379-263 & 146 & -3.0 & 41.5 \\
2M12214070+2707510 & 1\tablenotemark{1} & 55940.4279 & 5622-55940-237 & 91 & -5.40$\pm0.33$ & 6.29$\pm0.34$ \\
$\vert$ & 2\tablenotemark{1} & 55998.2603 & 5622-55998-238 & 115 & -11.23$\pm0.22$ & 12.32$\pm0.23$ \\
$\vert$ & 3\tablenotemark{1} & 56018.2352 & 5622-56018-274 & 129 & 11.63$\pm0.21$ & -10.55$\pm0.23$ \\
$\vert$ & 4\tablenotemark{1} & 56756.1755 & 7435-56756-273 & 138 & -3.96$\pm0.31$ & 5.32$\pm0.32$ \\
$\vert$ & 5\tablenotemark{1} & 56761.1460 & 7435-56761-279 & 119 & -3.20$\pm0.56$ & 3.72$\pm0.50$ \\
$\vert$ & 6\tablenotemark{1} & 56815.1664 & 7437-56815-274 & 132 & 22.10$\pm0.24$ & -20.30$\pm0.31$ \\
$\vert$ & 7\tablenotemark{1} & 56819.1401 & 7437-56819-279 & 100 & 20.23$\pm0.27$ & -18.62$\pm0.32$ \\
$\vert$ & 8\tablenotemark{1} & 56824.1421 & 7437-56824-279 & 123 & 17.59$\pm0.21$ & -16.07$\pm0.25$ \\
$\vert$ & 9\tablenotemark{1} & 56790.1321 & 7438-56790-238 & 118 & 16.55$\pm0.23$ & -15.74$\pm0.28$ \\
$\vert$ & 10\tablenotemark{1} & 56814.1431 & 7438-56814-274 & 136 & 22.40$\pm0.23$ & -20.25$\pm0.32$ \\
$\vert$ & 11\tablenotemark{1} & 56818.1447 & 7438-56818-274 & 132 & 20.75$\pm0.19$ & -19.17$\pm0.24$ \\
2M12260547+2644385 & 1\tablenotemark{1} & 55940.4279 & 5622-55940-255 & 285 & 6.94$\pm0.35$ & -8.34$\pm0.12$ \\
$\vert$ & 2\tablenotemark{1} & 55998.2603 & 5622-55998-291 & 349 & 11.85$\pm0.12$ & -13.72$\pm0.33$ \\
$\vert$ & 3\tablenotemark{1} & 56018.2352 & 5622-56018-160 & 467 & 11.00$\pm0.12$ & -14.29$\pm0.30$ \\
$\vert$ & 4\tablenotemark{1} & 56756.1755 & 7435-56756-154 & 436 & 8.99$\pm0.48$ & -9.06$\pm0.16$ \\
$\vert$ & 5\tablenotemark{1} & 56761.1460 & 7435-56761-195 & 393 & 6.96$\pm0.45$ & -10.20$\pm0.14$ \\
$\vert$ & 6\tablenotemark{1} & 56815.1664 & 7437-56815-245 & 422 & -16.14$\pm0.17$ & 22.71$\pm0.37$ \\
$\vert$ & 7\tablenotemark{1} & 56819.1401 & 7437-56819-206 & 340 & -14.77$\pm0.20$ & 20.49$\pm0.42$ \\
$\vert$ & 8\tablenotemark{1} & 56824.1421 & 7437-56824-290 & 393 & -12.06$\pm0.12$ & 15.93$\pm0.28$ \\
$\vert$ & 9\tablenotemark{1} & 56790.1321 & 7438-56790-195 & 370 & -17.14$\pm0.12$ & 22.87$\pm0.30$ \\
$\vert$ & 10\tablenotemark{1} & 56814.1431 & 7438-56814-208 & 454 & -16.86$\pm0.17$ & 23.68$\pm0.34$ \\
$\vert$ & 11\tablenotemark{1} & 56818.1447 & 7438-56818-201 & 477 & -15.20$\pm0.13$ & 20.70$\pm0.27$ \\
2M12260848+2439315 & 1 & 56756.1755 & 7435-56756-143 & 66 & -19.1 & 22.3 \\
$\vert$ & 2 & 56761.1460 & 7435-56761-131 & 62 & 18.7 & -12.5 \\
$\vert$ & 3 & 56815.1664 & 7437-56815-134 & 65 & -18.2 & 23.2 \\
$\vert$ & 4 & 56819.1401 & 7437-56819-137 & 43 & -14.5 & 21.5 \\
$\vert$ & 5 & 56824.1421 & 7437-56824-131 & 62 & 1.8 & 1.8 \\
$\vert$ & 6 & 56790.1321 & 7438-56790-137 & 55 & -13.3 & 19.8 \\
$\vert$ & 7 & 56814.1431 & 7438-56814-146 & 65 & -18.1 & 23.3 \\
$\vert$ & 8 & 56818.1447 & 7438-56818-131 & 73 & -16.2 & 23.2 \\
2M14545496+4108480 & 1 & 56378.3525 & 6852-56378-022 & 151 & 41.8 & -38.6 \\
$\vert$ & 2 & 56405.3445 & 6852-56405-010 & 123 & -22.4 & 32.9 \\
$\vert$ & 3 & 56409.3016 & 6852-56409-003 & 114 & 50.4 & -49.0 \\
$\vert$ & 4 & 56411.3084 & 6852-56411-022 & 152 & -19.0 & 29.8 \\
2M14551346+4128494 & 1 & 56378.3525 & 6852-56378-230 & 79 & -55.0 & 44.4 \\
$\vert$ & 2 & 56405.3445 & 6852-56405-218 & 74 & -31.9 & 15.6 \\
$\vert$ & 3 & 56409.3016 & 6852-56409-218 & 71 & 30.8 & -68.6 \\
$\vert$ & 4 & 56411.3084 & 6852-56411-221 & 93 & 26.6 & -66.7 \\
2M14562809+1648342 & 1 & 56432.2611 & 6844-56432-045 & 262 & 14.5 & 32.0 \\
$\vert$ & 2 & 56470.1510 & 6844-56470-033 & 279 & 16.8 & 2.7 \\
$\vert$ & 3 & 56703.4974 & 6844-56703-135 & 354 & -8.2 & 45.7 \\
2M17204248+4205070 & 1\tablenotemark{1} & 55992.4532 & 5670-55992-130 & 166 & 36.59$\pm0.30$ & -72.86$\pm0.61$ \\
$\vert$ & 2\tablenotemark{1} & 55998.4958 & 5671-55998-129 & 183 & 11.76$\pm0.37$ & -34.88$\pm0.66$ \\
$\vert$ & 3\tablenotemark{1} & 55999.4819 & 5670-55999-076 & 225 & 25.74$\pm0.36$ & -56.06$\pm0.89$ \\
$\vert$ & 4\tablenotemark{1} & 56019.3882 & 5670-56019-076 & 105 & 12.33$\pm0.52$ & -36.15$\pm0.91$ \\
$\vert$ & 5\tablenotemark{1} & 56022.4230 & 5670-56022-075 & 204 & 28.85$\pm0.33$ & -60.66$\pm0.84$ \\
$\vert$ & 6\tablenotemark{1} & 56023.4295 & 5670-56023-075 & 198 & -43.60$\pm0.33$ & 49.59$\pm0.68$ \\
$\vert$ & 7\tablenotemark{1} & 56025.4247 & 5671-56025-135 & 171 & 36.49$\pm0.56$ & -72.81$\pm1.79$ \\
$\vert$ & 8\tablenotemark{1} & 56027.4338 & 5671-56048-129 & 164 & -38.12$\pm0.32$ & 40.81$\pm0.88$ \\
$\vert$ & 9\tablenotemark{1} & 56048.3635 & 5671-56049-129 & 184 & 37.18$\pm0.30$ & -73.09$\pm0.63$ \\
$\vert$ & 10\tablenotemark{1} & 56049.3599 & 5837-56027-076 & 199 & -19.13$\pm0.30$ & 14.02$\pm0.77$ \\
$\vert$ & 11\tablenotemark{1} & 56054.3636 & 5838-56054-130 & 129 & 11.04$\pm0.43$ & -32.74$\pm0.82$ \\
$\vert$ & 12\tablenotemark{1} & 56084.2389 & 5838-56084-088 & 201 & 30.01$\pm0.32$ & -62.89$\pm0.80$ \\
$\vert$ & 13\tablenotemark{*} & 56429.2314 & - & 88 & 24.82$\pm0.16$ & -54.70$\pm0.22$ \\
$\vert$ & 14\tablenotemark{*} & 56434.2142 & - & 72 & -41.71$\pm0.16$ & 45.62$\pm0.21$ \\
$\vert$ & 15\tablenotemark{*} & 56439.2273 & - & 84 & 31.73$\pm0.16$ & -65.23$\pm0.20$ \\
$\vert$ & 16\tablenotemark{*} & 56444.2160 & - & 94 & -47.27$\pm0.16$ & 54.96$\pm0.18$ \\
$\vert$ & 17\tablenotemark{*} & 56472.3738 & - & 76 & 37.20$\pm0.16$ & -73.06$\pm0.20$ \\
$\vert$ & 18\tablenotemark{1} & 56052.4257 & 5837-56052-129 & 176 & -1.55$\pm0.52$ & -13.58$\pm1.07$ \\
$\vert$ & 19 & 56351.5046 & 5837-56351-076 & 176 & -5.2 & - \\
$\vert$ & 20 & 56362.5225 & 5838-56362-130 & 139 & -44.4 & 53.0 \\
2M19081153+2839105 & 1 & 56082.4258 & 5246-56082-274 & 144 & 33.1 & - \\
$\vert$ & 2 & 56086.4331 & 5246-56086-274 & 167 & 37.4 & 16.8 \\
$\vert$ & 3 & 56091.4578 & 5246-56091-279 & 95 & 38.1 & 17.3 \\
$\vert$ & 4 & 56084.4143 & 5247-56084-236 & 149 & 38.2 & 17.7 \\
$\vert$ & 5 & 56088.4135 & 5247-56088-281 & 133 & 33.0 & - \\
$\vert$ & 6 & 56092.4589 & 5247-56092-281 & 87 & 33.2 & - \\
$\vert$ & 7 & 56093.4469 & 5247-56093-281 & 121 & 37.4 & 16.7 \\
$\vert$ & 8 & 56779.3534 & 7475-56779-220 & 129 & 36.8 & 18.4 \\
$\vert$ & 9 & 56825.2622 & 7475-56825-273 & 111 & 23.3 & 33.7 \\
$\vert$ & 10 & 56831.3617 & 7476-56831-273 & 134 & 5.6 & 53.3 \\
$\vert$ & 11 & 56775.3457 & 7477-56775-274 & 81 & 28.4 & - \\
$\vert$ & 12 & 56781.3197 & 7477-56781-273 & 119 & 37.2 & 18.7 \\
$\vert$ & 13 & 56799.3399 & 7477-56799-280 & 131 & 37.7 & 16.4 \\
2M19235494+3834587 & 1\tablenotemark{1} & 55811.1100 & 5213-55811-283 & 41 & -52.70$\pm0.47$ & 12.32$\pm0.49$ \\
$\vert$ & 2\tablenotemark{1} & 55840.0915 & 5213-55840-241 & 50 & 1.68$\pm0.38$ & -41.62$\pm0.40$ \\
$\vert$ & 3\tablenotemark{1} & 55851.0793 & 5213-55851-283 & 58 & -1.49$\pm0.34$ & -38.29$\pm0.36$ \\
$\vert$ & 4\tablenotemark{*} & 56429.3213 & - & 35 & -38.99$\pm0.20$ & 0.00$\pm0.23$ \\
$\vert$ & 5\tablenotemark{*} & 56434.3099 & - & 34 & 40.99$\pm0.28$ & -83.73$\pm0.21$ \\
$\vert$ & 6\tablenotemark{*} & 56439.3063 & - & 33 & -89.67$\pm0.28$ & 46.22$\pm0.28$ \\
$\vert$ & 7\tablenotemark{*} & 56444.2909 & - & 45 & -27.41$\pm0.22$ & -10.09$\pm0.20$ \\
$\vert$ & 8\tablenotemark{*} & 56450.2800 & - & 35 & -100.80$\pm0.22$ & 60.46$\pm0.19$ \\
$\vert$ & 9\tablenotemark{*} & 56472.2149 & - & 38 & -50.90$\pm0.22$ & 12.27$\pm0.20$ \\
2M19433790+3225124 & 1 & 56603.0524 & 6079-56603-184 & 485 & -19.1 & - \\
$\vert$ & 2 & 56749.4836 & 7463-56749-231 & 470 & -18.5 & -24.5 \\
$\vert$ & 3 & 56754.4648 & 7463-56754-183 & 549 & 5.4 & -53.3 \\
2M20474087+3343054 & 1 & 56604.0570 & 6081-56604-160 & 247 & -46.7 & 15.5 \\
$\vert$ & 2 & 56611.0588 & 6081-56611-261 & 280 & 19.5 & -57.5 \\
$\vert$ & 3 & 56808.3481 & 7465-56808-159 & 236 & -18.3 & -18.5 \\
2M21005978+5103147 & 1 & 56623.0461 & 6949-56623-166 & 176 & -48.5 & -60.4 \\
$\vert$ & 2 & 56809.4544 & 6949-56809-165 & 127 & -83.5 & -26.4 \\
$\vert$ & 3 & 56626.0540 & 6950-56626-166 & 215 & -26.3 & -83.5 \\
$\vert$ & 4 & 56817.4235 & 7491-56817-165 & 206 & -68.9 & -40.3 \\
$\vert$ & 5 & 56836.4321 & 7493-56836-165 & 235 & -32.6 & -78.3 \\
2M21234344+4419277 & 1 & 56261.0499 & 6085-56261-117 & 83 & -164.5 & -77.8 \\
$\vert$ & 2 & 56479.4540 & 6085-56479-118 & 46 & -163.9 & -77.8 \\
$\vert$ & 3 & 56263.0643 & 6086-56263-117 & 85 & -86.6 & -162.2 \\
$\vert$ & 4 & 56485.4644 & 6086-56485-117 & 6 & -77.6 & -178.0 \\
$\vert$ & 5 & 56486.4616 & 6086-56486-117 & 30 & -105.8 & -141.5 \\
$\vert$ & 6 & 56589.1833 & 6086-56589-105 & 11 & -107.5 & -143.7 \\
$\vert$ & 7 & 56599.1220 & 6086-56599-105 & 45 & -65.2 & -186.3 \\
$\vert$ & 8 & 56603.1091 & 6086-56603-105 & 71 & -183.3 & -59.0 \\
2M21442066+4211363 & 1 & 55870.1160 & 5254-55870-046 & 59 & -70.6 & 53.6 \\
$\vert$ & 2 & 56170.2919 & 5254-56170-040 & 65 & -79.1 & 44.1 \\
$\vert$ & 3 & 56193.2333 & 5254-56193-046 & 80 & -79.3 & 45.0 \\
$\vert$ & 4 & 55869.1221 & 5255-55869-051 & 89 & 3.7 & -27.7 \\
$\vert$ & 5 & 56171.2829 & 5255-56171-087 & 79 & 10.1 & -47.9 \\
$\vert$ & 6 & 56204.1936 & 5255-56204-045 & 61 & 4.0 & -40.3 \\
$\vert$ & 7 & 56205.2364 & 6248-56205-089 & 103 & 26.5 & -67.7 \\
$\vert$ & 8 & 56221.2010 & 6248-56221-035 & 112 & 40.4 & -80.7 \\
$\vert$ & 9 & 56231.1930 & 6248-56231-035 & 82 & 44.2 & -80.1 \\
$\vert$ & 10 & 56206.2158 & 6249-56206-034 & 89 & -73.4 & 35.9 \\
$\vert$ & 11 & 56223.1801 & 6249-56223-034 & 87 & -76.6 & 41.6 \\
$\vert$ & 12 & 56232.1857 & 6249-56232-033 & 68 & -34.1 & -0.6 \\
\enddata
\tablecomments{Dashed out velocities indicate spurious RVs omitted from analysis. RVs not extracted via TODCor are assigned the ensemble uncertainty of $\sim 1.8 \frac{km}{s}$.}
\tablenotetext{*}{These visits are from the Hobby-Eberly Telescope (HET) High-Resolution Spectrograph (HRS). SNR is highly wavelength dependent for M-dwarfs; values reported here are at 7500 nm.  TODCOR was used for RV extraction. 
}
\tablenotetext{1}{TODCOR used for RV extraction.}
\tablenotetext{2}{Radial velocities for these epochs were mis-assigned by the extraction routine described in section \S3.1, and manually corrected.}
\tablecaption{}
\label{table:SB2_RVs}
\end{deluxetable*}
\fi

\clearpage

\setlength{\baselineskip}{0.6\baselineskip}
\bibliography{./references}
\setlength{\baselineskip}{1.667\baselineskip}
\end{document}